\documentclass[aps,rmp,nofootinbib]{revtex4}

\usepackage{hyperref}
\setcitestyle{numbers,square}
\usepackage{graphicx}

\begin{document}

\title{Generalized Dynamical Keldysh Model}

\author{E.Z. Kuchinskii, M.V.Sadovskii}

\affiliation{Institute for Electrophysics, Russian Academy of Sciences,
Ural Branch,\\ Amundsen str. 106, Ekaterinburg 620016, Russia}


\begin{abstract}
We consider a certain class of exactly solvable models, describing
spectral properties an electron moving in random in time external field with
different statistical characteristics. This electron can be band -- like or
belong to a quantum well. The known dynamical Keldysh model is generalized for
the case of fields with finite correlation time of fluctuations and for finite
transfer frequencies of these fluctuations. In all cases we are able to perform
the complete summation of all Feynman diagrams of corresponding perturbation
series for the Green's function. This can be done either by the reduction of
this series to some continuous fraction or by the use of the generalized Ward
identity from which we can derive recurrence relations for the Green's
function. In the case of a random field with finite transferred frequency
there appear the interesting effects of modulation of spectral density and
density of states.

{\sl Dedicated to 130-th anniversary of Pyotr Leonidovich Kapitza}

\end{abstract}

\maketitle

\section{Introduction}

While being an outstanding experimentalist, P.L. Kapitza sometimes addressed
also some purely theoretical problems. Well known is his elegant solution of
a problem of the motion of a classical particle in fast oscillating field
\cite{LL1}, where he essentially described this motion as a particle in a
random field with appropriate time averaging. Such fields and processes appear
in many problems of statistical radiophysics and radiotechnics, where a vast
literature exists \cite{Ry,Lev}. In quantum theory there is also multitude
problems of this kind.

In this work we shall consider a certain class of exactly solvable quantum
mechanical problems, related in general to the theory of electrons in
disordered systems and quantum structures, which is a dynamical generalization
of the so called Keldysh model.

The initial model was introduced by L.V. Keldysh in his unpublished thesis
in 1965 \cite{keld65}. Some of his results were used by A.L Efros in
Ref.  \cite{efros70}, devoted to doped semiconductors.
The detailed presentation of different aspects of this model in the general
context of electron theory of disordered systems was given in \cite{sad},
where the notion of ``Keldysh model'' was introduced for the first time.

In the following, the number of similar models were proposed, e.g. for the
description of the pseudogap appearing due to electron scattering by
fluctuations of short -- range order in one -- dimensional systems
\cite{sad,Sad1,Sad2,Won1,Won2,Sad3,SadTim}, which were later generalized for
two -- dimensional case to describe pseudogap in high -- temperature
superconductors \cite{PS1,PS2,SK98,ufn1,Tamm}.

Dynamical generalization of the initial Keldysh model for the case of electron
scattering by random in {\it time} fluctuations of external field was proposed
by Kikoin and Kiselev \cite{KK}, who considered electrons in quantum dots.
Detailed presentation of different results obtained for this and similar
models was given in Ref. \cite{EK}. The present paper is devoted to further
development and generalization of this type of models both for the case of
electrons in quantum dots and band -- like electrons in conductors of
different dimensionalities under the influence of dynamic random fields.

\section{Dynamical Keldysh model}

The model under consideration was proposed by Keldysh in 1965 \cite{keld65} as
some limiting case of problem of electron scattering by the random field of
static impurities in a disordered system \cite{sad,AGD}.
\begin{figure}[b]
\includegraphics[width=\textwidth]{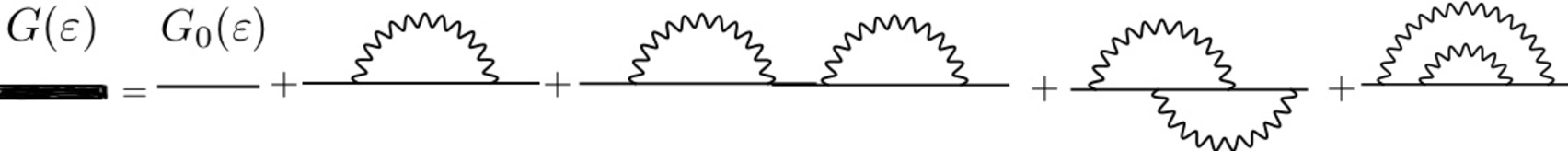}\\
\caption{Diagrammatic expansion for the Green's function. Double line
corresponds to ``dressed'' Green's function, wavy line corresponds to
correlator of Gaussian random field.}
\label{gf_pert}
\end{figure}
Keldysh has shown that the single -- particle Green's function in Gaussian
random field $V(r)$ with ``forward'' scattering (i.e. with zero
transferred momentum, corresponding to the limit of infinite spatial range of
fluctuations of the random potential) described by correlator
($d$ is spatial dimensionality):
\begin{equation}
D({\bf r}-{\bf r'})=\langle V({\bf r}) V({\bf r'})\rangle
= \Delta^2 \to D({\bf q}) =(2\pi)^d \Delta^2\delta({\bf q}),
\label{corpot}
\end{equation}
can be found by complete summation of all Feynman diagrams of perturbation
series. In fact, according to the usual diagram rules for the problem of
scattering by static random disorder \cite{sad,AGD}, diagram of $N$ - th
order contains $N$ interaction line with Gaussian random field
(denoted by by wavy lines), $2N+1$ solid lines, corresponding to Green's
functions and $2N$ vertices. The total number of diagrams in the given order of
perturbation theory $A_N$ corresponds to the total number of ways to connect
$2N$ vertices by $N$ interaction lines, which is equal to \cite{sad,sadk}:
\begin{equation}
A_N=(2N-1)!!= \frac{(2N-1)!}{2^{N-1} (N-1)!}.
\end{equation}
Diagrammatic contributions of the lowest orders in the series for single --
electron Green's function are shown in Fig. \ref{gf_pert}.
In this model all Feynman diagrams of the given order $N$ give the same
contributions to Green's function, so that the full series for it is of
the following form:
\begin{equation}
G(E)=G_0(E)\left\{1+\sum_{N=1}^\infty (2N-1)!! G_0^{2N}(E)
\Delta^{2N}\right\}.
\end{equation}
Further, to shorten notations we define  $E=\epsilon - \epsilon_{\bf p}$,
where $\epsilon_{\bf p}$ is free the electron spectrum, so that the ``bare''
Green's function is written as $G_0(E)=1/E$. Using integral representation of
$\Gamma$ -- function, we can use:
\begin{equation}
(2N-1)!! = \frac{1}{\sqrt{2\pi}}\int_{-\infty}^{\infty} d t t^{2N} e^{-t^2/2}
\end{equation}
so that the retarded Green's function (after the summation of geometric series)
can be written as:
\begin{equation}
G^R(E)=\frac{1}{\sqrt{2\pi \Delta^2}}\int_{-\infty}^{\infty}
d V\frac{e^{-V^2/2\Delta^2}}{E-V+i\delta}
\label{GrK}
\end{equation}
This equation has an obvious meaning \cite{sad} --- electron propagates
in spatially homogeneous Gaussian random field.
There is also another way to obtain this elegant result, which was also
proposed by Keldysh \cite{keld65} and later by Efros \cite{efros70}, and is
based on the use of an exact Ward identity, which allows the derivation of
differential equation for the Green's function. This equation has the
following form:
\begin{equation}
\Delta^2 \frac{d G(E)}{d E} + E\cdot G(E) =1.
\label{difeqG}
\end{equation}
Solving this equation with boundary condition $G(E\to\infty)=1/E$ immediately
leads to Eq. (\ref{GrK}) \cite{sad}.

Direct consequence of the obtained solution is the appearance of the Gaussian
``tail'' in the density of states of an electron in energy region
$\epsilon<0$ \cite{sad}.

In Refs. \cite{KK,EK} Keldysh model was reformulated for the case of electron
scattered by very slow {\em temporal} fluctuations of the random potential.
Appropriate dynamical Keldysh model can also be generalized for the case of
scattering by multiple component Gaussian non -- Markovian random fields \cite{EK}.

\begin{figure}[b]
\includegraphics[width=0.50\textwidth]{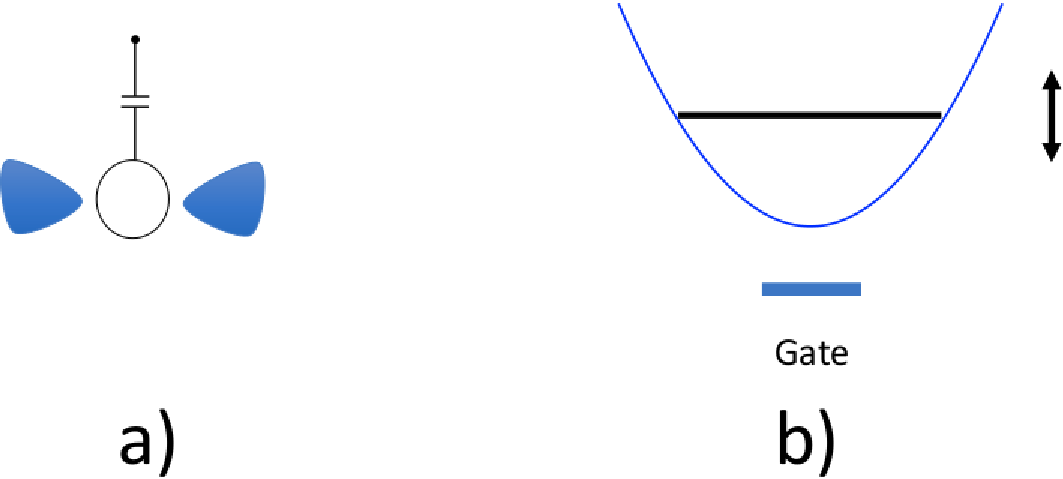}
\caption{(a) single quantum dot, with noise applied by external electrodes
(gate), (b) corresponding quantum well with fluctuating level.}
\label{sqd_1}
\end{figure}

As an example, following Refs. \cite{KK,EK} we may consider an electron in a
single quantum well (dot), which is formed by appropriate confining potential,
as shown in Fig. \ref{sqd_1}.
The gate creates external noise slowly changing confining potential of the well.

Single -- particle Hamiltonian for this problem has the following form:
\begin{equation}
H= \left[\epsilon_0+ V(t)\right] n.
\end{equation}
where $n=c^\dagger c$, and $c^\dagger$, $c$ are creation and annihilation
operators of an electron at the level within well. For simplicity we consider
spinless (spinpolarized) electrons. Classical potential random (Gaussian) in
time $V(t)$ is determined by its average value and pair correlation function:
\begin{equation}
\langle V(t)\rangle=0,\;\;\;\; \langle V(t) V(t')\rangle =D(t-t').
\label{V_t}
\end{equation}
For this function we assume the following form:
\begin{equation}
D(t-t')=\Delta^2 e^{-\gamma|t-t'|},
\label{Dcorr}
\end{equation}
where $\gamma =1/\tau$, with $\tau$ determining characteristic correlation time
of potential fluctuations, while $\Delta$ is the amplitude of the noise.
We may consider two limiting cases:
\begin{eqnarray}
\gamma &\to& \infty:\;\;\;\;  D(t-t')\to \Delta^2\delta(t-t'),\\
\gamma &\to& 0:\;\;\;\;  D(\omega)\to 2\pi \Delta^2\delta(\omega).
\label{Dlim}
\end{eqnarray}
Here $D(\omega)$ is the Fourier -- transform of $D(t-t')$.
The first case corresponds to ``fastest'' possible noise (``white'' noise) and
Markovian random process. The second case corresponds to slow noise,
with Keldysh model giving its slowest possible realization with (infinitely)
large relaxation time of fluctuations (infinite memory, of non -- Markovian
process).

Single -- electron (retarded) Green's function of electron in a well for the
given realization of the potential is:
\begin{equation}
G^R(\epsilon)=\frac{1}{\epsilon-\epsilon_0-V +i\delta}
\end{equation}
where $\epsilon_0$ is energy level in a well, while time -- averaging is again
reduced to Gaussian integration of this expression with distribution function
$P(V)=1/\sqrt{2\pi \Delta^2} \exp(-V^2/(2\Delta^2)$:
\begin{equation}
G^R(\epsilon)=\frac{1}{\sqrt{2\pi \Delta^2}}\int_{-\infty}^{\infty}
d V\frac{e^{-V^2/2\Delta^2}}{\epsilon-\epsilon_0-V+i\delta}
\label{GKex}
\end{equation}
Similarly we can consider an electron not within the well, but within energy
band of a system (placed between capacitor plates, on which a random noise is
generated) of any dimensionality. In this case it is just sufficient to make
a replacement $\epsilon_0\to\epsilon_{\bf p}$, where $\epsilon_{\bf p}$ is
band spectrum of an electron with quasimomentum ${\bf p}$.

The single -- well model is easily generalized also for the case of several
wells \cite{KK,EK}, which leads to Keldysh model with multicomponent noise.
Particularly interesting is the model of two quantum wells, which (in its
band -- like variant) is deeply related to an exactly solvable model of the
pseudogap state  \cite{Sad1,Sad2,Won1,Won2,Sad3,SadTim}. However, below we
shall only consider the single -- well model, leaving the two -- well case
(pseudogap fluctuations) for the separate work.

\section{Keldysh model and fluctuations with finite correlation time}

Below we show that an exact solution for the single -- particle Green's
function can also be obtained for Keldysh model with finite correlation time
of fluctuations $\tau=\gamma^{-1}$. This solution is easily found using the
method proposed by one of the authors in Ref. \cite{Sad3}, devoted to the model
of pseudogap in one -- dimensional systems.

\begin{figure}
\includegraphics[clip=true,width=0.5\textwidth,height=0.6\textheight]{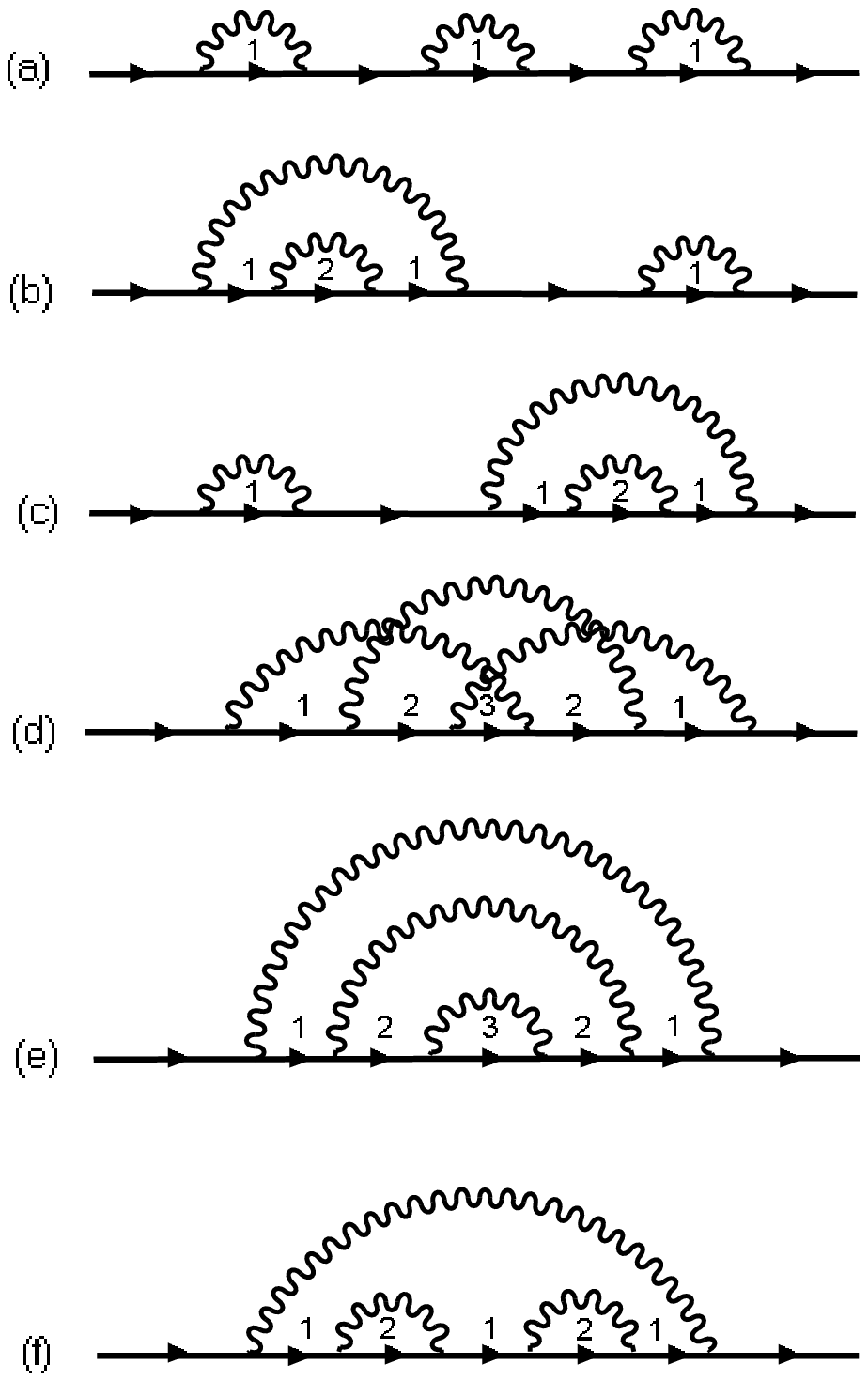}
\caption{Typical diagrams of the third order.}
\label{dia_3ord}
\end{figure}

Fourier -- transform of Eq. (\ref{Dcorr}), which is associated with interaction
lines in diagrams, can be written as:
\begin{equation}
D(\omega)=2\pi\Delta^2\frac{1}{\pi}\frac{\gamma}{\omega^2+\gamma^2}=
2\pi\Delta^2\frac{1}{\pi}\frac{\gamma}{(\omega+i\gamma)(\omega-i\gamma)}
\label{corrf}
\end{equation}
For $\gamma\to 0$ this is naturally reduced to the second expression in
(\ref{Dlim}). Let us clarify the calculations of a diagram of an an arbitrary
order. In fact this can be done exactly. As an example let us consider some
typical diagrams of third order shown in Fig. \ref{dia_3ord}.
We can easily calculate the contribution of an arbitrary diagram as we can
actually guarantee that nonzero contribution to integrals (over transferred
frequencies) appear only from the poles of Lorentzians\footnote{In the problem
analyzed in Ref. \cite{Sad3} this statement is only approximate \cite{sadk}.
Here all calculations (frequency integrations) are performed exactly.}
$D(\omega)$. For example, elementary calculations show, that contribution
of diagram in Fig. \ref{dia_3ord} (d) to the retarded Green's function has the
following form:
\begin{eqnarray}
\Delta^6\frac{1}{\epsilon-\epsilon_0}\frac{1}{\epsilon-\epsilon_0
+i\gamma}\frac{1}{\epsilon-\epsilon_0+2i\gamma}\frac{1}{\epsilon-
\epsilon_0+3i\gamma}
\frac{1}{\epsilon-\epsilon_0+2i\gamma}\frac{1}{\epsilon-
\epsilon_0+i\gamma}\frac{1}{\epsilon-\epsilon_0}
\label{3ordr}
\end{eqnarray}
Contributions of arbitrary diagrams are quite similar: integers  $k$, written
above electronic lines Fig. \ref{dia_3ord}, show have many times the term
$i\gamma$ enters corresponding denominator. Note that contribution of
diagram with crossing interaction lines in Fig. \ref{dia_3ord} (d) are just
equal to the contribution of diagram with no intersections of interaction lines
shown in Fig.  \ref{dia_3ord} (e). This is a manifestation of the general
property -- contribution of any diagram with crossing interaction lines is
equal to the contribution of some diagram with no intersections \cite{Sad3}.
Precisely because of this property we can introduce an exact algorithm of
complete summation of Feynman series.

Details of combinatorics and rules to reduce diagrams with crossing interaction
lines to those without intersections were considered in Ref. \cite{Sad3}
(see also Ref. \cite{sad})\footnote{In the problem under consideration here
combinatorics of diagrams is reduced to commensurate case of Ref. \cite{Sad3}.}.
One can easily convince himself that the number of irreducible diagrams for
self -- energy which are equal to the given diagram with no intersections of
interaction lines is equal to the product of certain combinatorial factors
$v(k)$ ($k$ is the number of $i\gamma$ contributions in the denominator of
the Green's function in diagram without intersections, standing below
$k$ interaction lines) which are associated with consequent interaction lines
of this diagram.
Correspondingly in the following we can use just the diagrams with no intersections
of interaction lines associating extra combinatorial factors $v(k)$ to
interaction lines of such diagrams. In our case $v(k)=k$ \cite{Sad3}.

\begin{figure}
\includegraphics[clip=true,width=0.5\textwidth]{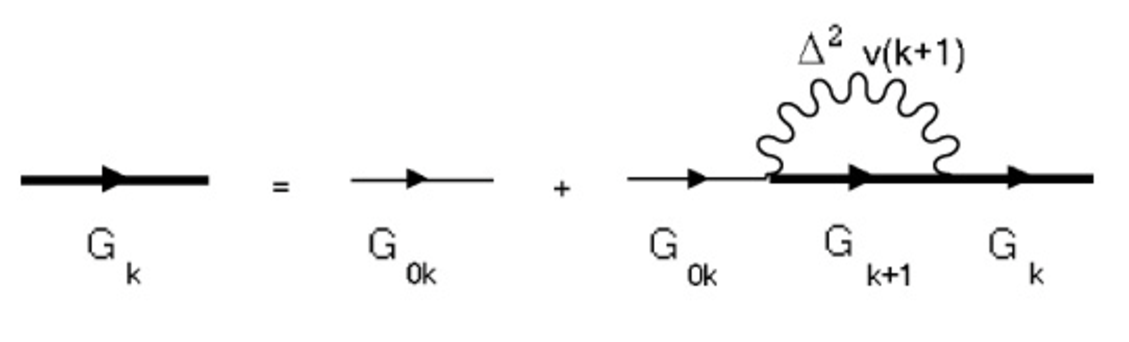}
\caption{``Dyson equation'' representation of recurrence equation for the
Green's function. Here we introduced $G_{0k}=[\epsilon-\epsilon_0+ik\gamma]^{-1}$.}
\label{dia_Dyson}
\end{figure}

Then we can easily obtain the recursion relation determining the irreducible
self -- energy, which includes {\em all} diagrams of corresponding Feynman
series \cite{sad,Sad3}:
\begin{equation}
\Sigma_k(\epsilon,\epsilon_0)=\frac{\Delta^2v(k)}{\epsilon
-\epsilon_0+ik\gamma-\Sigma_{k+1}(\epsilon,\epsilon_0)};\ v(k)=k
\label{Si_rec}
\end{equation}
Then we immediately get the recursion relation for Green's function itself:
\begin{equation}
G_k(\epsilon,\epsilon_0)=\{\epsilon-\epsilon_0+ik\gamma-
\Delta^2v(k+1)G_{k+1}(\epsilon,\epsilon_0)\}^{-1},
\label{G_rec}
\end{equation}
and the {\em physical} Green's function is defined as $G(\epsilon,\epsilon_0)
\equiv G_{k=0}(\epsilon,\epsilon_0)$, which is equivalent the complete
sum of Feynman series for our model. In fact these relations give the following
{\em continuous -- fraction} representation of single -- electron Green's function:
\begin{eqnarray}
G(\epsilon,\epsilon_0)
=\frac{1}{\displaystyle \epsilon-\epsilon_0-
\frac{\Delta^2}{\displaystyle \epsilon-\epsilon_0+i\gamma-
\frac{2\Delta^2}{\displaystyle \epsilon-\epsilon_0+
2i\gamma-\frac{3\Delta^2}{\displaystyle \epsilon-
\epsilon_0+3i\gamma-...}}}}
\label{G_chain}
\end{eqnarray}
Symbolically our recursion relation can be represented as a kind of
``Dyson equation'', shown in Fig. \ref{dia_Dyson}.

For $\gamma=0$ we can use the following continuous -- fraction representation of
incomplete (upper) $\Gamma$ -- function:
\begin{equation}
\Gamma(\alpha,x)=\int_x^{\infty}dte^{-t}t^{\alpha-1}=\frac{x^{\alpha}}
{x+\frac{1-\alpha}{1+\frac{1}{x+\frac{2-\alpha}{1+...}}}}
\end{equation}
to convince ourselves that Eq. (\ref{G_chain}) reproduces an exact result of
(\ref{GKex}) obtained by direct summation of all diagrams.

\section{Fluctuations with finite transferred frequency and finite correlation time}

Let us consider now more general case of fluctuations with finite characteristic
frequency $\omega_0$. We shall again consider classical potential random in
time $V(t)$ (\ref{V_t}) with pair correlation function:

\begin{eqnarray}
D(t-t')=\Delta^2 e^{-\gamma|t-t'|}\cos[\omega_0(t-t')]
=\frac{\Delta^2}{2}e^{-\gamma|t-t'|}
\left[
e^{i\omega_0(t-t')}+e^{-i\omega_0(t-t')}
\right].
\label{Dcorr_w0}
\end{eqnarray}
For $\omega_0=0$ we obtain again correlator  (\ref{Dcorr}) and the model
with zero transferred frequency considered above.

\begin{figure}
\includegraphics[clip=true,width=0.5\textwidth]{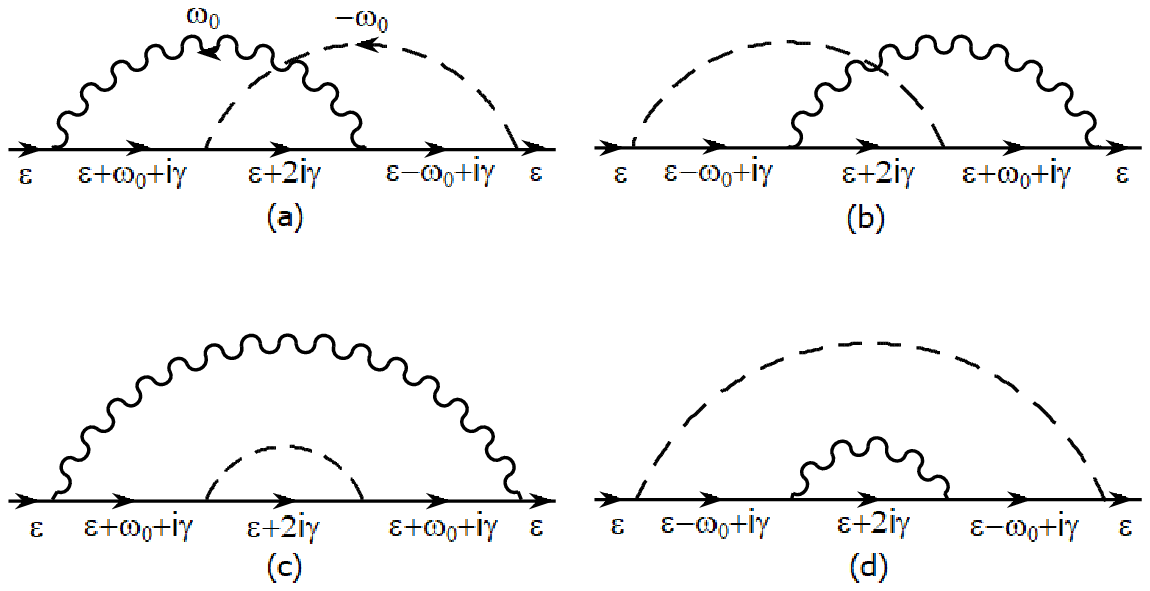}
\caption{Typical diagrams of second order.}
\label{dia_2ord}
\end{figure}

Fourier -- transform of correlator (\ref{Dcorr_w0}) has the form:
\begin{equation}
D(\omega)=2\pi\frac{\Delta^2}{2}
\left[
\frac{1}{\pi}\frac{\gamma}{(\omega -\omega_0)^2+\gamma^2}+
\frac{1}{\pi}\frac{\gamma}{(\omega +\omega_0)^2+\gamma^2}
\right].
\label{corrf_w0}
\end{equation}
Thus in corresponding diagram technique we have two sorts of interaction lines --
wavy and dashed, transferring frequencies $+\omega_0$ and $-\omega_0$
correspondingly. Both interaction lines lead to addition of $i\gamma$ term to
energy $\epsilon$ in each electron Green's function, which is below corresponding
interaction line. In Fig. \ref{dia_2ord} we show typical second order diagrams.
It is easy to see that in current model the contribution of diagrams with
intersecting interaction lines does not necessarily coincide with some
diagram without such intersections. However, we still can obtain an exact
solution for the single -- electron Green's function using the generalized
Ward identity.

\subsection{Generalized Ward identity and recurrence equations for the Green's function}

\begin{figure}
\includegraphics[clip=true,width=0.5\textwidth]{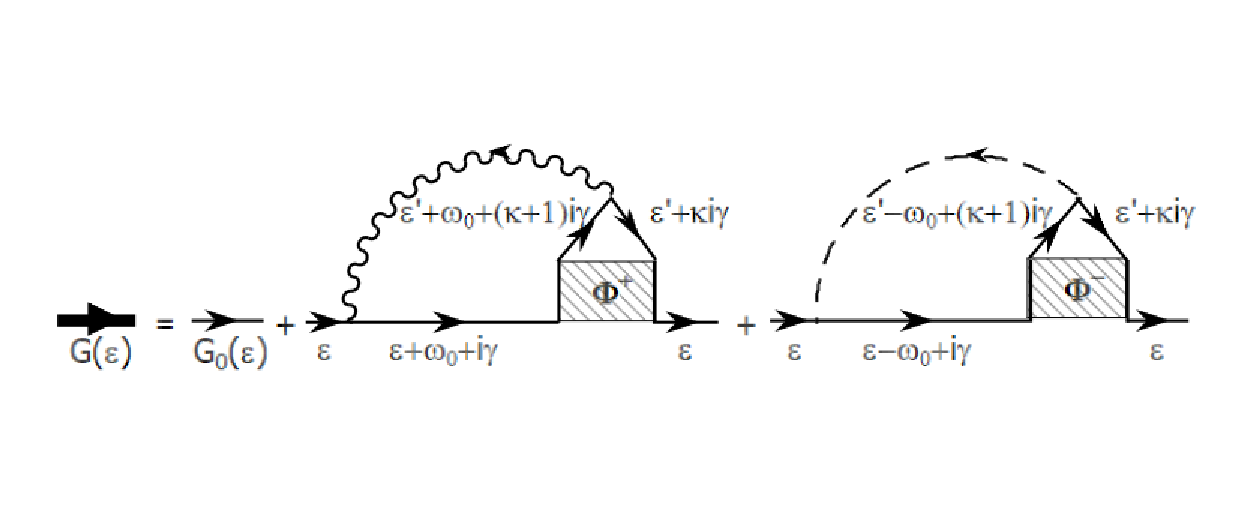}
\caption{Diagrammatic representation of equation for the Green's function}
\label{dia_eqG}
\end{figure}

Single -- electron Green's function $G$ can be easily determined via the
full two -- particle function $\Phi$:
\begin{equation}
G(\epsilon)=G_0(\epsilon)+
G_0(\epsilon)\frac{\Delta^2}{2}
\left\{
\sum_{\epsilon '}\Phi_{\epsilon\epsilon '}(\omega_0)+
\sum_{\epsilon '}\Phi_{\epsilon\epsilon '}(-\omega_0)
\right\}.
\label{G_Phi}
\end{equation}
Here $\Phi$ is the full two -- particle Green's function, including four
external electronic lines and contribution corresponding to the product of two
``dressed'' single -- particle Green's functions $G$.
To shorten expressions in our analysis we make a replacement
$\epsilon - \epsilon_0 \to \epsilon$, i.e. count energies from energy level
in the well $\epsilon_0$, then $G_0(\epsilon)=1/\epsilon$.
Diagrammatic representation of Eq. (\ref{G_Phi}) for the Green's function
is shown in Fig. \ref{dia_eqG}.
To find two -- particle Green's functions $\Phi$ entering Eq. (\ref{G_Phi})
we shall use the generalized Ward identity \cite{PRB07}, which in this
purely dynamical model takes the following form:
\begin{equation}
G(\epsilon+\omega)-G(\epsilon)=-\sum_{\epsilon '}\Phi_{\epsilon\epsilon '}(\omega)
\left\{
G_0^{-1}(\epsilon '+\omega)-G_0^{-1}(\epsilon ')
\right\}.
\label{TW_full}
\end{equation}
Here the expression in figure brackets in the r.h.s.
$G_0^{-1}(\epsilon '+\omega)-G_0^{-1}(\epsilon ')=\epsilon '+\omega -\epsilon '=\omega$ is independent of $\epsilon '$,
so that we immediately obtain:
\begin{equation}
\sum_{\epsilon '}\Phi_{\epsilon\epsilon '}(\omega)=-\frac{G(\epsilon+\omega)-G(\epsilon)}{\omega}.
\label{Phi_w}
\end{equation}
In the current problem any interaction line again adds $i\gamma$ term to energy of
electronic lines below it, i.e. effectively our interaction lines transfer a
complex frequency $\pm\omega_0+i\gamma$. Then Ward identity (\ref{TW_full})
for the vertex with $+\omega_0$ takes the form:
\begin{eqnarray}
G(\epsilon+\omega_0+i\gamma)-G(\epsilon)
=-\sum_{\epsilon '}\Phi_{\epsilon\epsilon '}(\omega_0)
(\epsilon '+\omega_0+(k+1)i\gamma -(\epsilon '+ki\gamma))
=-(\omega_0+i\gamma)\sum_{\epsilon '}\Phi_{\epsilon\epsilon '}(\omega_0).
\label{TW_+w0}
\end{eqnarray}
As a result for the two -- particle Green's function with $+\omega_0$ vertex
we obtain:
\begin{equation}
\sum_{\epsilon '}\Phi_{\epsilon\epsilon '}(\omega_0)=
-\frac{G(\epsilon+\omega_0+i\gamma)-G(\epsilon)}{\omega_0+i\gamma}.
\label{Phi_+w0}
\end{equation}
Similarly for $\Phi$ with $-\omega_0$ vertex we get:
\begin{equation}
\sum_{\epsilon '}\Phi_{\epsilon\epsilon '}(-\omega_0)=
-\frac{G(\epsilon-\omega_0+i\gamma)-G(\epsilon)}{-\omega_0+i\gamma}.
\label{Phi_-w0}
\end{equation}
Substituting these two -- particle functions (\ref{Phi_+w0}) and
(\ref{Phi_-w0}) into Eq. (\ref{G_Phi}), we obtain the functional equation
for the Green's function:
\begin{eqnarray}
G(\epsilon)=G_0(\epsilon)-
G_0(\epsilon)\frac{\Delta^2}{2}
\left\{
\frac{G(\epsilon+\omega_0+i\gamma)-G(\epsilon)}{\omega_0+i\gamma}+
\frac{G(\epsilon-\omega_0+i\gamma)-G(\epsilon)}{-\omega_0+i\gamma}
\right\}
\label{eqG_0}
\end{eqnarray}
so that:
\begin{equation}
G(\epsilon)=\frac{1-\frac{\Delta^2}{2}
\left[\frac{G(\epsilon+\omega_0+i\gamma)}{\omega_0+i\gamma}+
\frac{G(\epsilon-\omega_0+i\gamma)}{-\omega_0+i\gamma}\right]}
{G_0^{-1}(\epsilon)+\Delta^2\frac{i\gamma}{\omega_0^2+\gamma^2}}.
\label{eqG_fin}
\end{equation}
It should be noted that the use of the generalized Ward identity (\ref{TW_full})
allows also an exact solution (reducing to the integral equation) of the
problem of finding the single -- particle Green's function
$G(\epsilon)$ OED an electron in random {\it Gaussian} potential with
arbitrary correlator $D(\omega)$. Equation for the Green's function in this
case has the following form:
\begin{equation}
G(\epsilon)=G_0(\epsilon)+
G_0(\epsilon)\int_{-\infty}^{+\infty}\frac{d\omega}{2\pi}D(\omega)
\sum_{\epsilon '}\Phi_{\epsilon\epsilon '}(\omega).
\label{G_DPhi}
\end{equation}
Using Ward identity (\ref{TW_full}) we immediately obtain (\ref{Phi_w}) and
the integral equation for the Green's function:
\begin{equation}
G(\epsilon)=G_0(\epsilon)-
G_0(\epsilon)\int_{-\infty}^{+\infty}\frac{d\omega}{2\pi}D(\omega)
\frac{G(\epsilon+\omega)-G(\epsilon)}{\omega}.
\label{eqG_D}
\end{equation}
If we use $D(\omega)$ in the form given by Eq. (\ref{corrf_w0}) the frequency
integral here is easily calculated. The second factor in the integrand does not
contain pole at $\omega =0$ and is analytic in the upper half -- plane of
complex  $\omega$, so that closing the integration contour above,
we obtain the  contribution to integral only from the poles at
$\omega =\pm\omega_0+i\gamma$ of two Lorentzians in (\ref{corrf_w0}) immediately
getting (\ref{eqG_0}), and functional equation (\ref{eqG_fin}).

Solving Eq. (\ref{eqG_fin}) by iterations, starting from initial the
approximation
\begin{equation}
\widetilde{G}_0(\epsilon)=\frac{1}{G_0^{-1}(\epsilon)+\Delta^2\frac{i\gamma}{\omega_0^2+\gamma^2}},
\label{tildeG_0}
\end{equation}
one can easily see that each iteration adds to energy (besides $\pm\omega_0$)
additional $i\gamma$ term. Thus we can introduce the following notations:
\begin{equation}
G_n(\epsilon)\equiv G(\epsilon+ni\gamma) \qquad G_{0n}(\epsilon)\equiv G_0(\epsilon+ni\gamma)=
\frac{1}{\epsilon+ni\gamma},
\label{GnG0n}
\end{equation}
where $n=0,1,2\ldots$ and apply Eq. (\ref{eqG_fin}) for energy
$\epsilon+ni\gamma$, making replacement $\epsilon\to\epsilon+ni\gamma$.
Then in notations of (\ref{GnG0n}) equation (\ref{eqG_fin}) takes the
form\footnote{Naturally, Eq. (\ref{recG_fin}) can be also obtained directly
using the generalized Ward identity applying it for energy
$\epsilon+ni\gamma$.}:
\begin{equation}
G_n(\epsilon)=\frac{1-\frac{\Delta^2}{2}
\left[\frac{G_{n+1}(\epsilon+\omega_0)}{\omega_0+i\gamma}+
\frac{G_{n+1}(\epsilon-\omega_0)}{-\omega_0+i\gamma}\right]}
{G_{0n}^{-1}(\epsilon)+\Delta^2\frac{i\gamma}{\omega_0^2+\gamma^2}}.
\label{recG_fin}
\end{equation}
As a result we obtain the recursion procedure where at each ``storey'' $n$
$G_n$ depends only on real energy. Numerical realization of such procedure is
rather simple. At some high ``storey'' $n=N\gg 1$ we define a set of
$G_N(\epsilon)$, e.g. $G_N(\epsilon)=0$.
Then, withe the help of  (\ref{recG_fin}) and interpolation we find the set
$G_{N-1}(\epsilon)$ etc., until we reach the physical
$G(\epsilon)=G_{n=0}(\epsilon)$.

For $\omega_0=0$ we return to the model with zero transferred frequency
and finite correlation time described above. In this limit the recursion
equation (\ref{recG_fin}) takes the form:
\begin{equation}
G_n(\epsilon)=\frac{1+i\frac{\Delta^2}{\gamma}G_{n+1}(\epsilon)}
{G_{0n}^{-1}(\epsilon)+i\frac{\Delta^2}{\gamma}}.
\label{recG_w00}
\end{equation}
Visually the recursion procedure (\ref{recG_w00}) has nothing in common
with procedure (\ref{G_rec}), leading to continuous -- fraction representation
of $G$ given by Eq.  (\ref{G_chain}). However, direct numerical calculations show
that both produce absolutely same results for the physical Green's function
$G_{n=0}(\epsilon)$ (in the limit of initial ``storey'' $N\to\infty$).

For $\gamma =0$ in the limit of $\omega_0\to 0$ Eq. (\ref{eqG_0}) immediately
reduces to differential equation (\ref{difeqG}) for the Green's function in the
usual Keldysh model, as
$\lim\limits_{\omega_0\to 0}\frac{G(\epsilon+\omega_0)-G(\epsilon)}{\omega_0}=
\lim\limits_{\omega_0\to 0}\frac{G(\epsilon-\omega_0)-G(\epsilon)}{-\omega_0}=
\frac{dG(\epsilon)}{d\epsilon}$.
Green's function $G(\epsilon)$ is analytic in the upper half -- plane of complex
energy $\epsilon$ and the derivative $\frac{dG(\epsilon)}{d\epsilon}$ gives the
same result along different directions of $d\epsilon$ in this half -- plane.
Thus for other order of limits $\omega_0=0$, $\gamma\to 0$ from
Eq. (\ref{eqG_0}) we again obtain the differential equation (\ref{difeqG}).
Analyticity of the Green's function allows to write it (in the upper half --
plane of $\epsilon$) as:
\begin{equation}
G(\epsilon)=\int_{-\infty}^{\infty}d\epsilon '\frac{\rho(\epsilon ')}{\epsilon-\epsilon '},
\label{analG}
\end{equation}
where  $\rho(\epsilon)=-\frac{1}{\pi}ImG(\epsilon)$ is the spectral density
(density of states for the quantum dot). Then in this limit in Eq. (\ref{eqG_0})
we get:
\begin{eqnarray}
&&\lim_{\gamma\to 0}\frac{G(\epsilon+i\gamma)-G(\epsilon)}{i\gamma}=\nonumber\\
&&=\lim_{\gamma\to 0}\frac{1}{i\gamma}\int_{-\infty}^{\infty}d\epsilon '\rho(\epsilon ')
\left[\frac{1}{\epsilon+i\gamma -\epsilon '}-\frac{1}{\epsilon -\epsilon '}\right]
=\nonumber\\
&&=-\int_{-\infty}^{\infty}d\epsilon '\frac{\rho(\epsilon ')}{(\epsilon -\epsilon ')^2}=
\frac{dG(\epsilon)}{d\epsilon}
\label{dG_gam}
\end{eqnarray}
Analytic properties of Green's function (\ref{analG}) allow to reduce the
functional equation (\ref{eqG_fin}) to integral equation for spectral density
$\rho(\epsilon)$. Let us rewrite functional equation (\ref{eqG_fin}) as:
\begin{eqnarray}
G(\epsilon)=\widetilde{G}_0(\epsilon)-\widetilde{G}_0(\epsilon)\frac{\Delta^2}{2}
\left[\frac{G(\epsilon+\omega_0+i\gamma)}{\omega_0+i\gamma}+
\frac{G(\epsilon-\omega_0+i\gamma)}{-\omega_0+i\gamma}\right],
\label{eqG1_fin}
\end{eqnarray}
where $\widetilde{G}_0(\epsilon)$, defined in (\ref{tildeG_0}), can be written
as:
\begin{equation}
\widetilde{G}_0(\epsilon)=\frac{1}{\epsilon+i\Gamma}.
\label{G0_tilde}
\end{equation}
Here
\begin{equation}
\Gamma=\frac{\Delta^2\gamma}{\omega_0^2+\gamma^2}
\label{Gamma}
\end{equation}
is an effective non -- perturbative damping due to the random field.
Then for the spectral density we immediately obtain:
\begin{eqnarray}
\rho(\epsilon)=\widetilde{\rho}_0(\epsilon)+\frac{\Delta^2}{2\pi}
Im\left\{\widetilde{G}_0(\epsilon)\left[\frac{G(\epsilon+\omega_0+i\gamma)}{\omega_0+i\gamma}+
\frac{G(\epsilon-\omega_0+i\gamma)}{-\omega_0+i\gamma}\right]\right\}
\label{eqDOS_int}
\end{eqnarray}
where $\widetilde{\rho}_0(\epsilon)=-\frac{1}{\pi}Im\widetilde{G}_0(\epsilon)=
\frac{1}{\pi}\frac{\Gamma}{\epsilon^2+\Gamma^2}$ is an effective ``bare''
spectral density (density of states).
Eq. (\ref{eqDOS_int}) is easily solved numerically by iterations, starting
from initial approximation $\rho(\epsilon)=\widetilde{\rho}_0(\epsilon)$.

\subsection{Exact solution for the Green's function in the form of infinite series}

Eq. (\ref{eqG1_fin}) can be solved by iterations starting from
$\widetilde{G}_0(\epsilon)$.
If we represent the result of each iteration as simple fractions
(so that there are no $\epsilon$ in the coefficients), one can easily convince
himself, that the Green's function $G$ becomes the sum of
$\widetilde{G}_0(\epsilon+(n-m)\omega_0+(n+m)i\gamma)$, where $n$ and $m$
are integers, with coefficients independent of $\epsilon$.
Thus we look for the solution for the Green's function in the following form:
\begin{equation}
G(\epsilon)=\sum_{n,m=0}^\infty A_{nm}
\frac{1}{\epsilon+(n-m)\omega_0+(n+m)i\gamma+i\Gamma},
\label{G_sumA}
\end{equation}
where coefficients $A_{nm}$ are independent of $\epsilon$ and can be found
substituting (\ref{G_sumA}) into (\ref{eqG1_fin}).
Then we have:
\small
\begin{eqnarray}
&&\widetilde{G}_0(\epsilon)G(\epsilon+\omega_0+i\gamma)=\nonumber\\
&&=\sum\limits_{n,m=0}^\infty A_{nm}\frac{1}{\epsilon+i\Gamma}
\frac{1}{\epsilon+i\Gamma+(n+1-m)\omega_0+(n+1+m)i\gamma}=\nonumber\\
&&=\sum\limits_{n,m=0}^\infty A_{nm}\frac{1}{(n+1)(\omega_0+i\gamma)+m(-\omega_0+i\gamma)}\times\nonumber\\
&&\times\left[\frac{1}{\epsilon+i\Gamma}-
\frac{1}{\epsilon+i\Gamma+(n+1)(\omega_0+i\gamma)+m(-\omega_0+i\gamma)}\right]
\label{GG_+w0}
\end{eqnarray}
\begin{eqnarray}
&&\widetilde{G}_0(\epsilon)G(\epsilon-\omega_0+i\gamma)=\nonumber\\
&&=\sum\limits_{n,m=0}^\infty A_{nm}\frac{1}{n(\omega_0+i\gamma)+(m+1)(-\omega_0+i\gamma)}\times\nonumber\\
&&\times\left[\frac{1}{\epsilon+i\Gamma}-
\frac{1}{\epsilon+i\Gamma+n(\omega_0+i\gamma)+(m+1)(-\omega_0+i\gamma)}\right]
\label{GG_-w0}
\end{eqnarray}
\normalsize
Substituting (\ref{GG_+w0}), (\ref{GG_-w0}) into (\ref{eqG1_fin}) we find the
coefficient $A_{00}$ before $\frac{1}{\epsilon+i\Gamma}$ as:
\small
\begin{eqnarray}
&&A_{00}=1-\frac{\Delta^2}{2}\times\nonumber\\
&&\times\left[\frac{1}{\omega_0+i\gamma}
\sum\limits_{n,m=0}^\infty A_{nm}\frac{1}{(n+1)(\omega_0+i\gamma)+m(-\omega_0+i\gamma)}+\right.\nonumber\\
&&\left.+\frac{1}{-\omega_0+i\gamma}
\sum\limits_{n,m=0}^\infty A_{nm}\frac{1}{n(\omega_0+i\gamma)+(m+1)(-\omega_0+i\gamma)}
\right].
\label{A00}
\end{eqnarray}
\normalsize
For other coefficients:
\begin{eqnarray}
A_{nm}=\frac{\Delta^2}{2}\frac{1}{n(\omega_0+i\gamma)+m(-\omega_0+i\gamma)}
\left[\frac{A_{n-1m}}{\omega_0+i\gamma}+\frac{A_{nm-1}}{-\omega_0+i\gamma}\right].
\label{Anm}
\end{eqnarray}
Naturally we have $A_{-1m}=A_{n-1}=0$.

Eq. (\ref{Anm}) allows to obtain the whole set of coefficients at
$n_f=n+m$ ``storey'' from the values of coefficients at  $n_f-1$ ``storey'',
and finally to express all coefficients via  $A_{00}$.
Coefficients obtained for several lower ``storeys'' allow us to guess, that
the general form of the coefficients can be written as:
\begin{equation}
A_{nm}=\frac{A_{00}}{n!m!}\left(\frac{\Delta^2}{2}\right)^{n+m}\frac{1}{(\omega_0+i\gamma)^{2n}(-\omega_0+i\gamma)^{2m}}.
\label{Anm_fin}
\end{equation}
Substitution of $A_{nm}$ from (\ref{Anm_fin}) into Eq. (\ref{Anm})
confirms this guess.

Now using Eq. (\ref{A00}) we can find $A_{00}$:
\small
\begin{eqnarray}
&&A_{00}=1-\frac{\Delta^2}{2}\times\nonumber\\
&&\times\left[\frac{1}{\omega_0+i\gamma}
\sum\limits_{n=1,m=0}^\infty A_{n-1m}\frac{1}{n(\omega_0+i\gamma)+m(-\omega_0+i\gamma)}+\right.\nonumber\\
&&\left.+\frac{1}{-\omega_0+i\gamma}
\sum\limits_{n=0,m=1}^\infty A_{nm-1}\frac{1}{n(\omega_0+i\gamma)+m(-\omega_0+i\gamma)}
\right].\nonumber\\
\label{A00_1}
\end{eqnarray}
\normalsize
Using (\ref{Anm}) in (\ref{Anm_fin}) we get:
\begin{eqnarray}
&&A_{00}=1-\sum\limits_{
{n,m}{(n+m\ne0)}}A_{nm}=\nonumber\\
&&=1-\sum\limits_{
{n,m}{(n+m\ne0)}}
\frac{A_{00}}{n!m!}\left(\frac{\Delta^2}{2}\right)^{n+m}
\frac{1}{(\omega_0+i\gamma)^{2n}(-\omega_0+i\gamma)^{2m}}.\nonumber\\
\label{A00_2}
\end{eqnarray}
Finally $A_{00}$ takes the form:
\begin{eqnarray}
&&A_{00}=\frac{1}
{\sum\limits_{n=0}^\infty\frac{1}{n!}\left(\frac{\Delta^2}{2}\right)^n
\frac{1}{(\omega_0+i\gamma)^{2n}}
\sum\limits_{m=0}^\infty\frac{1}{m!}\left(\frac{\Delta^2}{2}\right)^m
\frac{1}{(-\omega_0+i\gamma)^{2m}}}=\nonumber\\
&&=e^{-\frac{\Delta^2}{2(\omega_0+i\gamma)^2}}e^{-\frac{\Delta^2}{2(\omega_0+i\gamma)^2}}=
e^{-\frac{\Delta^2(\omega_0^2-\gamma^2)}{(\omega_0^2+\gamma^2)^2}}
\label{A00_fin}
\end{eqnarray}
As a result we obtain the following expression for the Green's function
(\ref{G_sumA}):
\begin{eqnarray}
G(\epsilon)=e^{-\frac{\Delta^2(\omega_0^2-\gamma^2)}{(\omega_0^2+\gamma^2)^2}}
\sum_{n,m=0}^\infty\frac{1}{n!}\frac{1}{m!}
\frac{1}{(\omega_0+i\gamma)^{2n}(-\omega_0+i\gamma)^{2m}}
\left(\frac{\Delta^2}{2}\right)^{n+m}
\frac{1}{\epsilon+(n-m)\omega_0+(n+m)i\gamma+i\Gamma}
\label{G_sum_fin}
\end{eqnarray}

Let us briefly analyze the limiting behavior of the Green's function and
corresponding spectral density $\rho(\epsilon)=-\frac{1}{\pi}ImG(\epsilon)$
following from (\ref{G_sum_fin}).

In the limit of $\gamma\to 0$ we get:
\begin{equation}
G(\epsilon)=e^{-\frac{\Delta^2}{\omega_0^2}}
\sum_{n,m=0}^\infty\frac{1}{n!}\frac{1}{m!}
\left(\frac{\Delta^2}{2\omega_0^2}\right)^{n+m}
\frac{1}{\epsilon+(n-m)\omega_0+i\delta },
\label{G_sum_gam0}
\end{equation}
and spectral density has the form:
\begin{equation}
\rho(\epsilon)=e^{-\frac{\Delta^2}{\omega_0^2}}
\sum_{n,m=0}^\infty\frac{1}{n!}\frac{1}{m!}
\left(\frac{\Delta^2}{2\omega_0^2}\right)^{n+m}
\delta(\epsilon+(n-m)\omega_0)
\label{N_sum_gam0}
\end{equation}
which is the set of $\delta$ peaks at $\epsilon =\pm k\omega_0$.
The weights of these peaks (coefficients before corresponding $\delta$ --
functions) are:
\begin{eqnarray}
S^{(+k)}=S^{(-k)}=e^{-\frac{\Delta^2}{\omega_0^2}}
\sum_{n=0}^\infty\frac{1}{n!(n+k)!}
\left(\frac{\Delta^2}{2\omega_0^2}\right)^{2n+k}=
e^{-\frac{\Delta^2}{\omega_0^2}}
I_k\left(\frac{\Delta^2}{\omega_0^2}\right),
\label{Sk_gam0}
\end{eqnarray}
where $I_k$ -- is the modified Bessel function of imaginary argument.
The total area of all these peaks is:
\begin{eqnarray}
&&S=\sum_{k=-\infty}^\infty S^{(k)}=e^{-\frac{\Delta^2}{\omega_0^2}}
\sum_{n,m=0}^\infty\frac{1}{n!}\frac{1}{m!}
\left(\frac{\Delta^2}{2\omega_0^2}\right)^{n+m}=\nonumber\\
&&=e^{-\frac{\Delta^2}{\omega_0^2}}
\sum_{n=0}^\infty\frac{1}{n!}\left(\frac{\Delta^2}{2\omega_0^2}\right)^{n}
\sum_{m=0}^\infty\frac{1}{m!}\left(\frac{\Delta^2}{2\omega_0^2}\right)^{m}=1,
\label{Ssum_gam0}
\end{eqnarray}
as it should be.

In the limit of $\omega_0\to 0$ we return to the model of fluctuations with
finite correlation time and from Eq. (\ref{G_sum_fin}) we obtain:
\begin{eqnarray}
&&G(\epsilon)=e^{\frac{\Delta^2}{\gamma^2}}
\sum_{n,m=0}^\infty\frac{1}{n!}\frac{1}{m!}
\left(-\frac{\Delta^2}{2\gamma^2}\right)^{n+m}
\frac{1}{\epsilon+(n+m)i\gamma+i\frac{\Delta^2}{\gamma}}=\nonumber\\
&&=e^{\frac{\Delta^2}{\gamma^2}}
\sum_{k=0}^\infty
\left[\sum_{n=0}^k\frac{1}{n!(k-n)!}\right]
\left(-\frac{\Delta^2}{2\gamma^2}\right)^k
\frac{1}{\epsilon+ki\gamma+i\frac{\Delta^2}{\gamma}}.
\label{G1_sum_w00}
\end{eqnarray}
As $\sum\limits_{n=0}^k\frac{k!}{n!(k-n)!}=2^k$ we get for the Green's function:
\begin{equation}
G(\epsilon)=e^{\frac{\Delta^2}{\gamma^2}}
\sum_{k=0}^\infty\frac{1}{k!}
\left(-\frac{\Delta^2}{\gamma^2}\right)^k
\frac{1}{\epsilon+ki\gamma+i\frac{\Delta^2}{\gamma}}=
\label{G_sum_w00}
\end{equation}
\begin{equation}
=e^{\frac{\Delta^2}{\gamma^2}}\frac{1}{i\gamma}
\left(\frac{\Delta^2}{\gamma^2}\right)^
{-\left(\frac{\epsilon}{i\gamma}+\frac{\Delta^2}{\gamma^2}\right)}
\gamma\left(\frac{\epsilon}{i\gamma}+\frac{\Delta^2}{\gamma^2},\frac{\Delta^2}{\gamma^2}\right),
\label{G_int_w00}
\end{equation}
where
\begin{equation}
\gamma(\alpha,x)=\int_0^{x}dte^{-t}t^{\alpha-1}
\label{gam_sml}
\end{equation}
is incomplete (lower) $\Gamma$ -- function.
Eqs. (\ref{G_sum_w00}) and (\ref{G_int_w00}) can be considered as series and
integral representations for continuous fraction of (\ref{G_chain}).

The problem of an electron in Gaussian field of dynamic fluctuations with finite
correlation time has much in common with the problem of Holstein polaron in
semiconductors with low mobility, i.e. with the problem of finding the
single electron Green's function in Holstein 
model \cite{Holstein} of an
electron interacting with optical phonon mode with frequency $\Omega$ in the
limit of transfer integral between nearest neighbors $t\to 0$ ($t\ll\Omega$).
Usually such problem is analyzed by making Lang -- Firsov canonical
transformation \cite{LangFirsov} in Holstein Hamiltonian \cite{Holstein}
However, the diagram technique for electron -- phonon interaction in this
model is completely equivalent to diagram technique in our model of dynamical
fluctuations with finite correlation time after the replacement:
\begin{equation}
\Delta\to g  \qquad i\gamma\to -\Omega
\label{replacement}
\end{equation}
where $g$ is electron -- phonon coupling constant. We only have to take into
account that in this diagram technique in the denominators of electron Green's
functions we have continuous addition $-\Omega$ terms instead of
$i\gamma$, because of two terms in phonon propagator:
\begin{equation}
D(\omega)=\frac{1}{\omega-\Omega+i\delta}-\frac{1}{\omega+\Omega-i\delta}
\label{Ph_D}
\end{equation}
only the first term contribute to frequency integrals due to the fact that all
electronic Green's functions in this problem are retarded.

Thus the Green's function of Holstein polaron (for $t\to 0$) is determined by
continuous fraction (\ref{G_chain}) with replacement (\ref{replacement}).
For the first time Holstein polaron Green's function of this form was derived
in Ref. \cite{Berciu06}. Our series expression for the Green's function
(\ref{G_sum_w00}) in the model of dynamical fluctuations with finite correlation
time immediately allows us to get (after the replacement (\ref{replacement}))
the well known exact result for the Green's function of Holstein polaron as
\cite{LangFirsov,Berciu06}:
\begin{equation}
G(\epsilon)=e^{-\frac{g^2}{\Omega^2}}
\sum_{k=0}^\infty\frac{1}{k!}
\left(\frac{g^2}{\Omega^2}\right)^k
\frac{1}{\epsilon-k\Omega+\frac{g^2}{\Omega}+i\delta}=
\label{G_LF}
\end{equation}
Note that our use of the Ward identity is in some sense equivalent to
Lang -- Firsov transformation in Holstein polaron problem. An effective ``bare''
Green's function (\ref{G0_tilde})with non -- perturbative damping (\ref{Gamma}),
appearing due to the use of the Ward identity, in the model with
$\omega_0=0$ is:
\begin{equation}
\widetilde{G}_0(\epsilon)=\frac{1}{\epsilon+i\frac{\Delta^2}{\gamma}}.
\label{G0_tilde_w00}
\end{equation}
which in the Holstein polaron problem, after the replacement (\ref{replacement}),
takes the form:
\begin{equation}
\widetilde{G}_0(\epsilon)=\frac{1}{\epsilon+\frac{g^2}{\Omega}+i\delta},
\label{G0_tilde_LF}
\end{equation}
appearing after Lang -- Firsov transformation of an effective ``bare''
Green's function of polaron with non -- perturbative shift of the ground state
$\epsilon_0=-\frac{g^2}{\Omega}$
\cite{LangFirsov,Berciu06}.

\section{Numerical results}

Now for the most general model of fluctuations with finite frequency and
correlation time we actually have three exact numerical procedures to find
the Green's function:
\begin{enumerate}
\item recursive procedure (\ref{recG_fin}),
\item integral equation for spectral density (\ref{eqDOS_int}),
\item series representation (\ref{G_sum_fin}).
\end{enumerate}
For the wide range of parameters ($\Delta$, $\gamma$, $\omega_0$) of the model
our numerical calculations showed that all three procedures lead to absolutely
same results for spectral density (density of states).
Of these, the recursion procedure (\ref{recG_fin}) is most fast for numerics,
though for small values of $\gamma\ll\Delta,\omega_0$ and $\omega_0<0.3\Delta$
it requires significant increase of the number of energies in corresponding array
and the number of an initial ``storey'' to start, while series representation
(\ref{G_sum_fin}) in this range of parameters is well convergent.
However, the series representation is inappropriate for direct numerical
analysis in the region of $\Delta\gg\gamma >\omega_0$, which is connected
both with large values of the exponent before the series and with the large
number of terms in the series to be taken into account to compensate this
exponent.

\begin{figure}
\includegraphics[clip=true,width=0.5\textwidth]{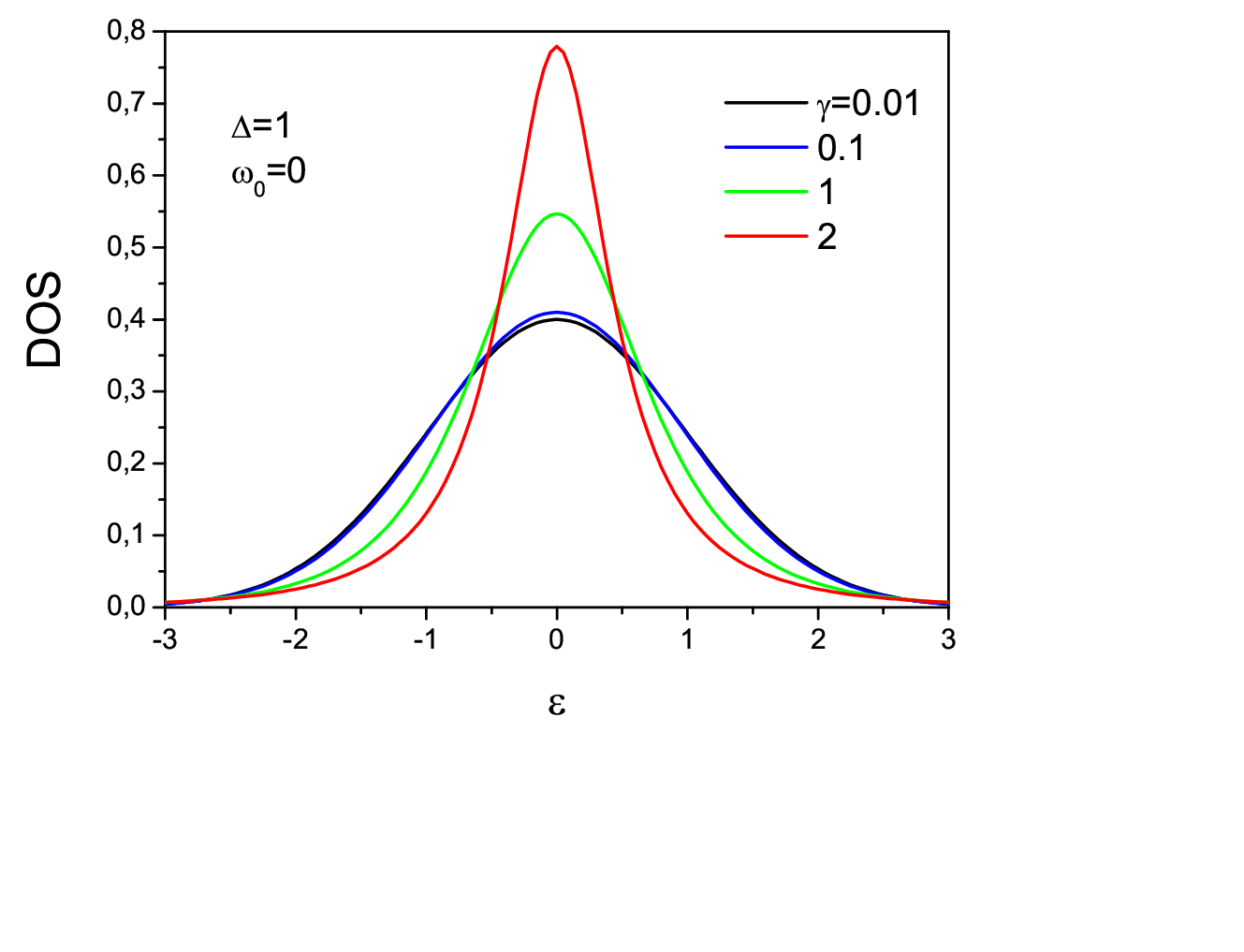}
\caption{Spectral density (density of states in quantum dot) in the model with
finite correlation time ($\omega_0=0$) for different values of $\gamma$.}
\label{DOSw00_gam}
\end{figure}

Now let us discuss our numerical results.
In Fig.\ref{DOSw00_gam} we demonstrate evolution of the spectral density
with increasing $\gamma$ (i.e. with decreasing correlation time of fluctuations)
for the model with $\omega_0=0$. For $\gamma=0$ (in the usual Keldysh model)
spectral density is Gaussian with the width $\Delta$ (dispersion -- $\Delta^2$).
The growth of $\gamma$ leads to decrease of characteristic width of the spectral
density with appropriate growth of $\rho(0)$.

\begin{figure}
\includegraphics[clip=true,width=0.40\textwidth]{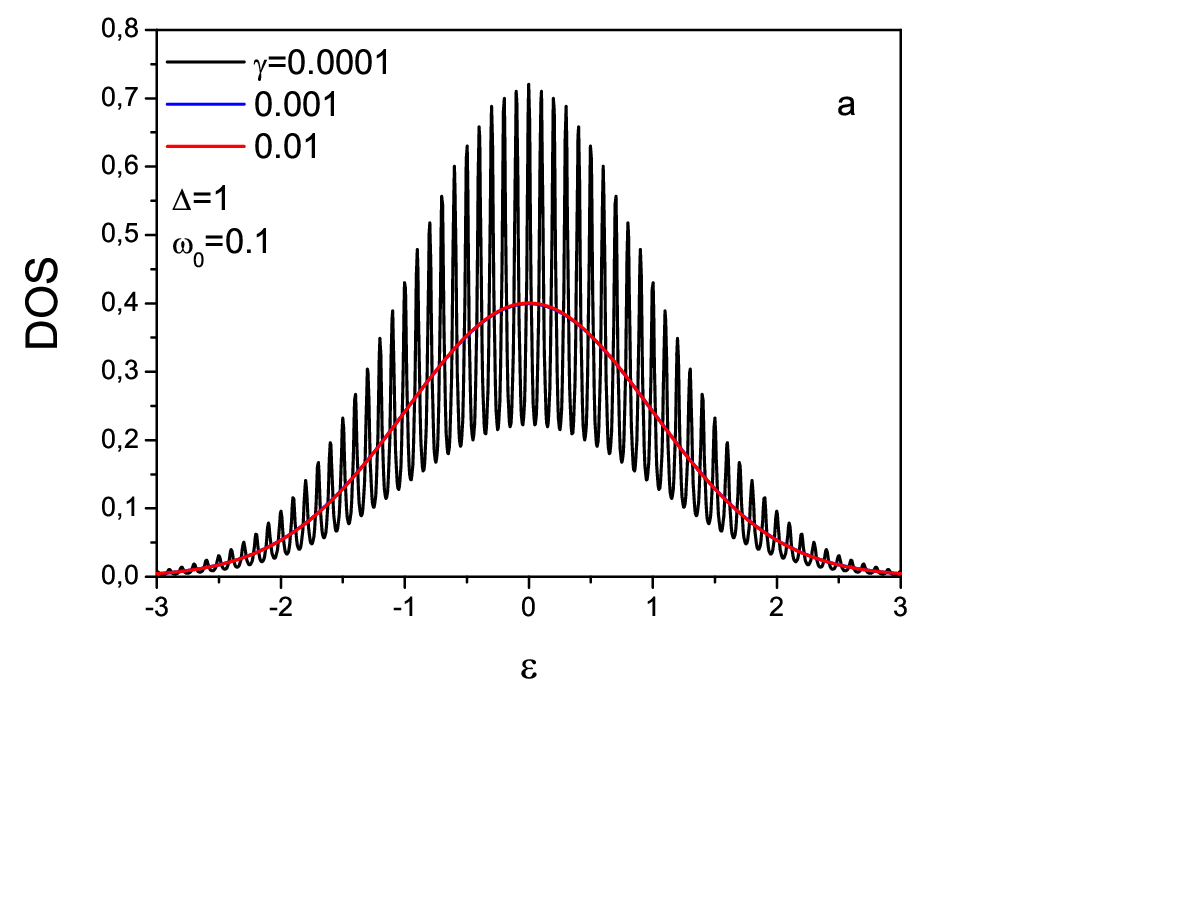}
\includegraphics[clip=true,width=0.40\textwidth]{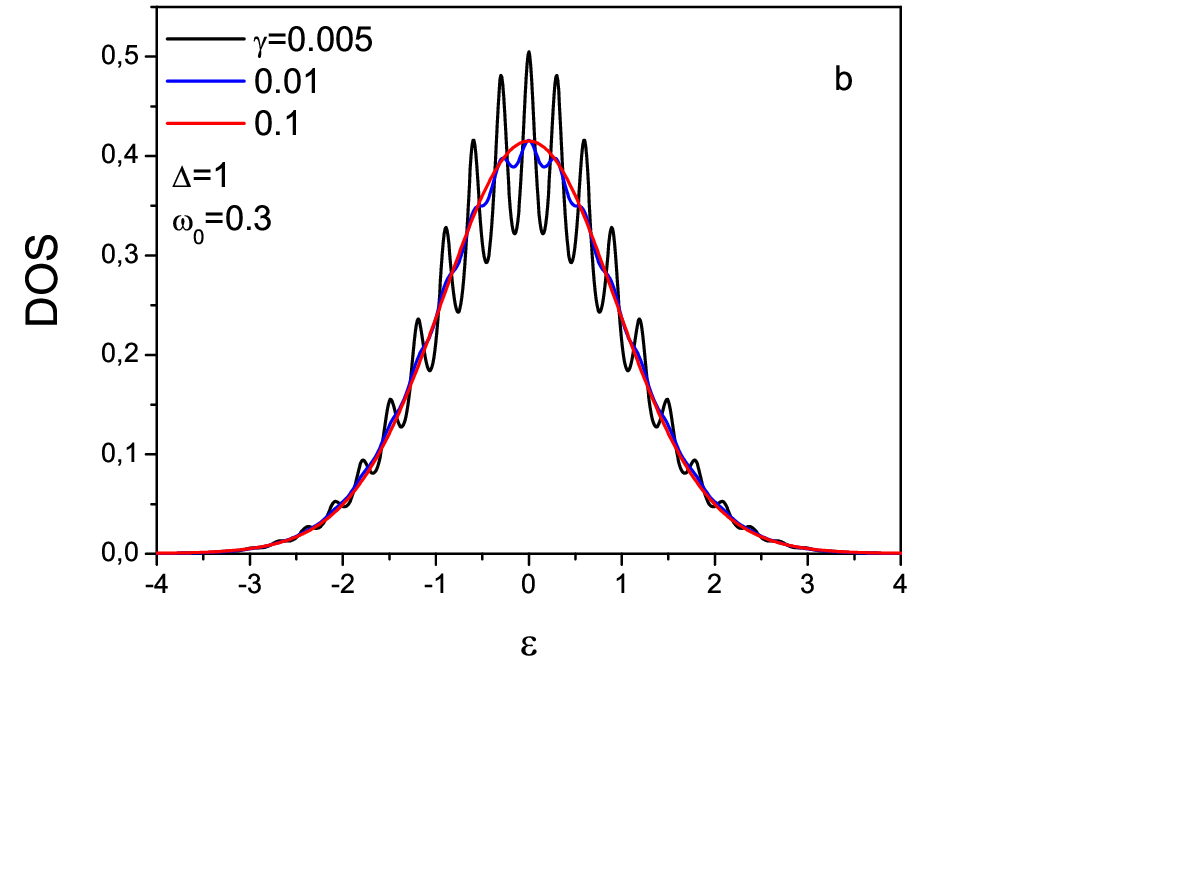}
\includegraphics[clip=true,width=0.40\textwidth]{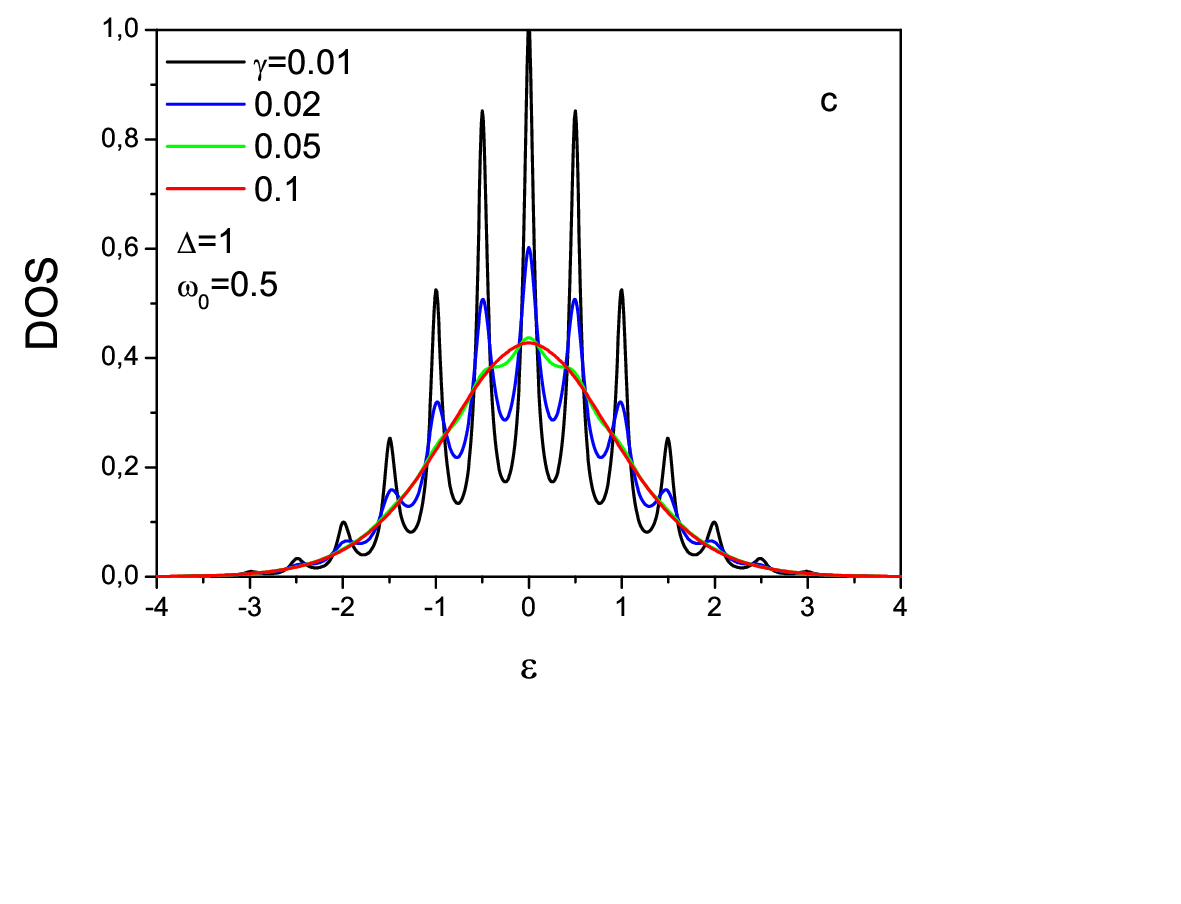}
\includegraphics[clip=true,width=0.40\textwidth]{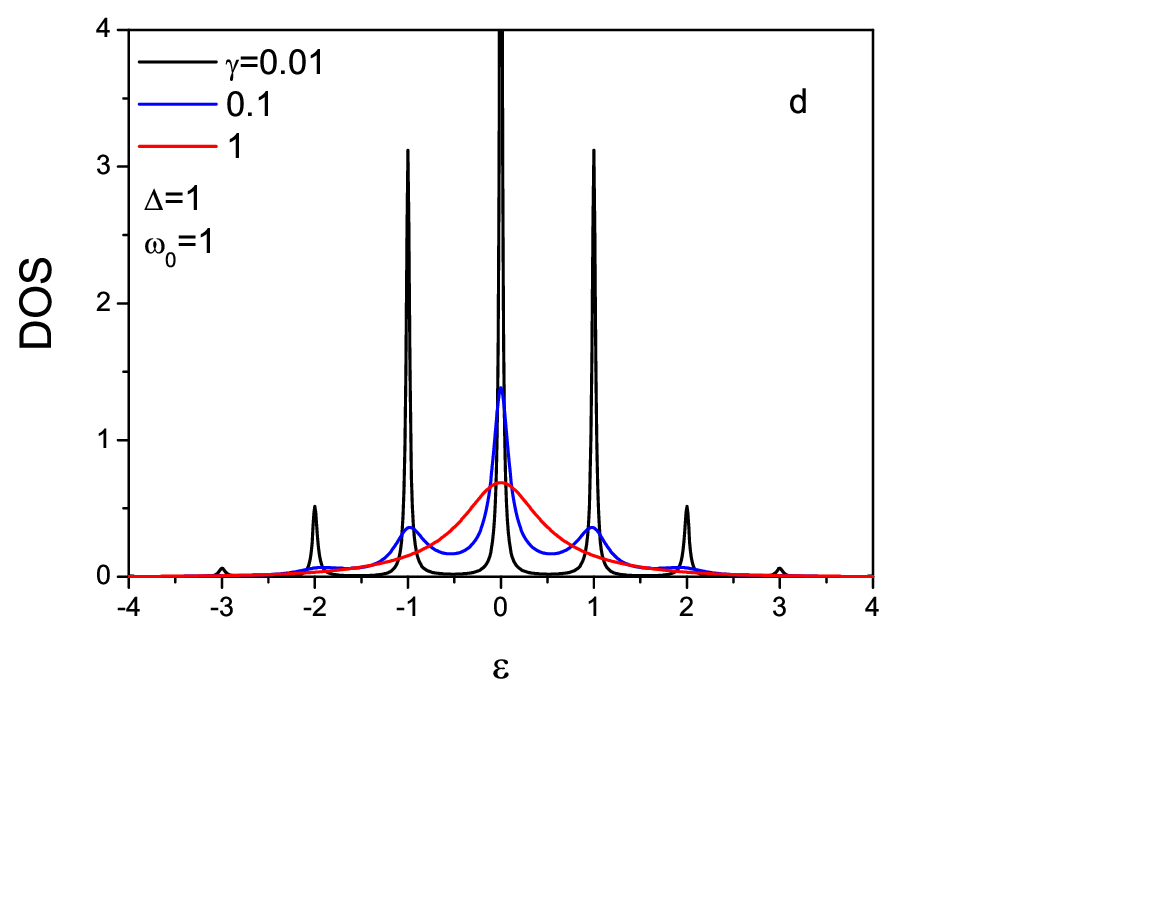}
\includegraphics[clip=true,width=0.40\textwidth]{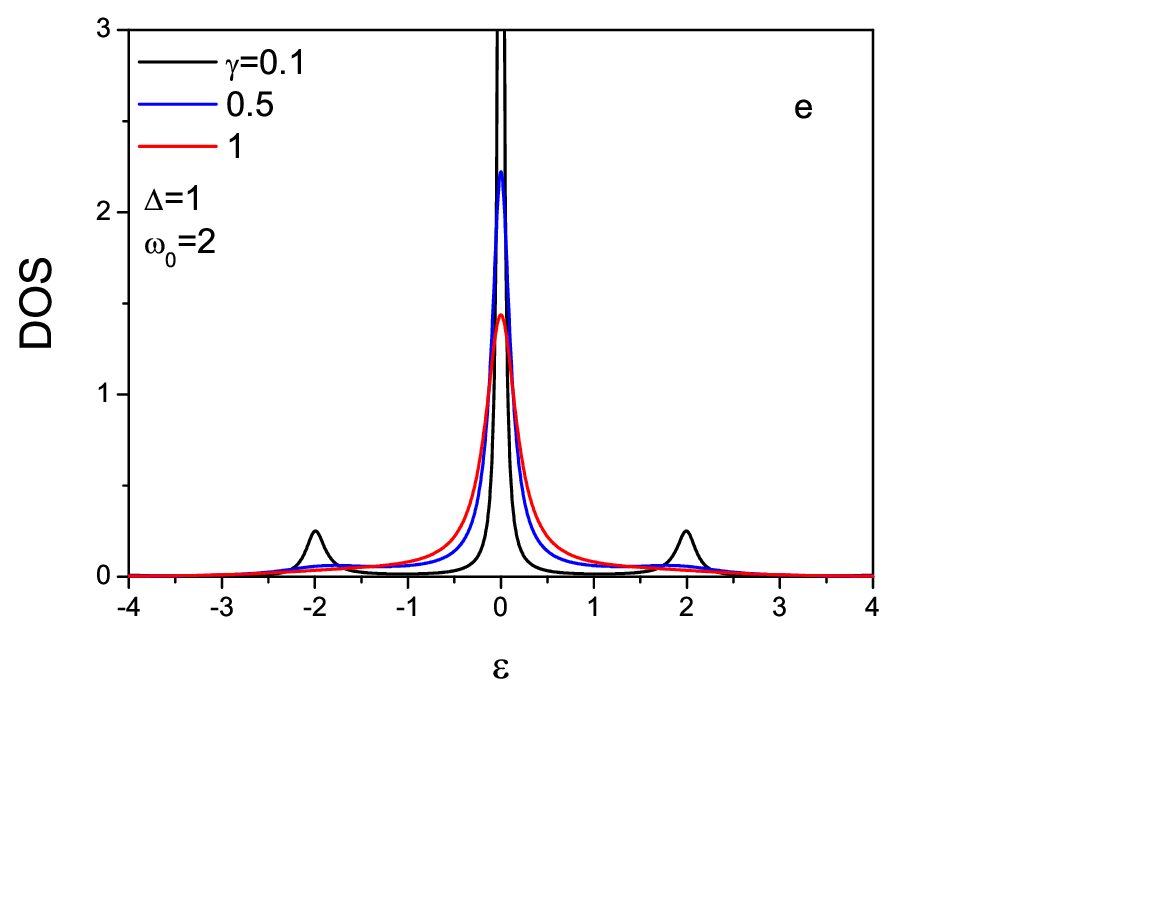}
\includegraphics[clip=true,width=0.40\textwidth]{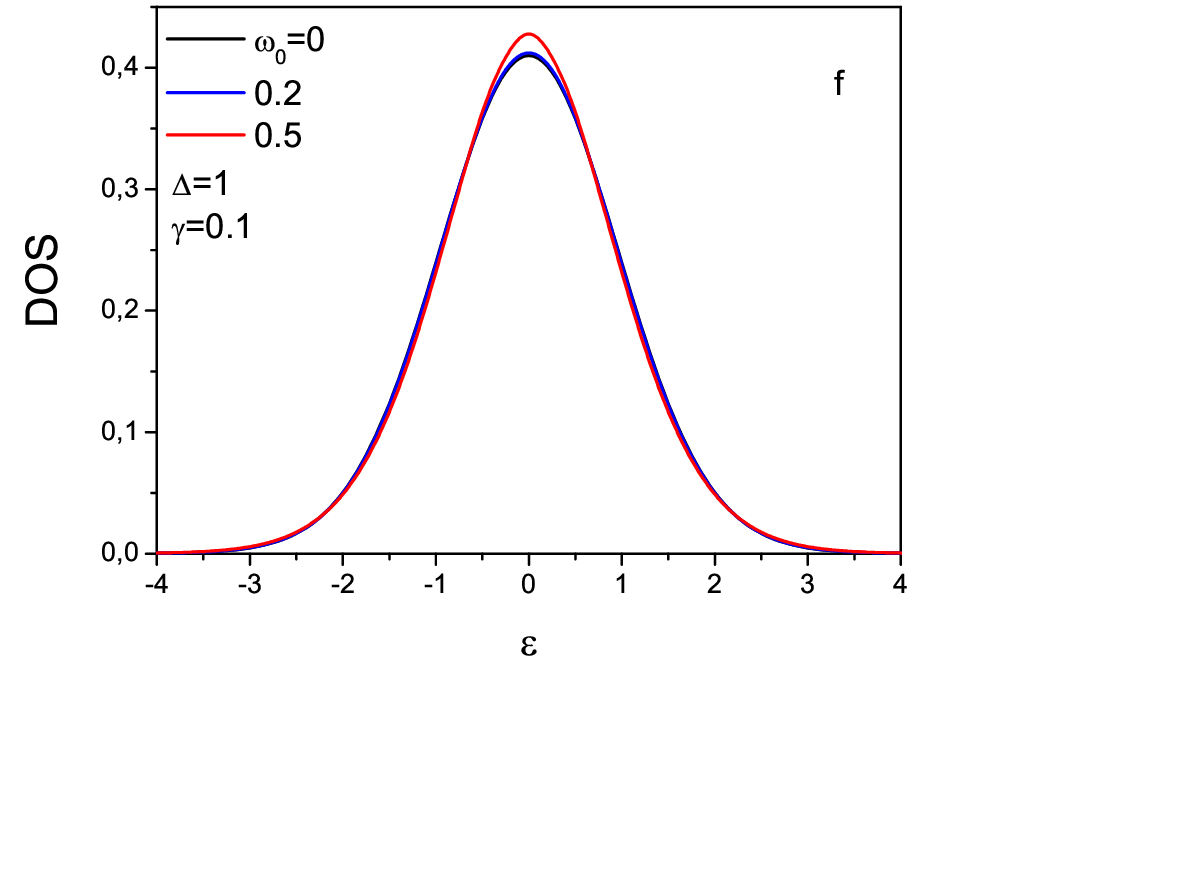}
\caption{Spectral density (density of states) of the quantum dot in the model
with finite transfer frequency and relaxation time for $\Delta=1$ and different
values of $\omega_0$ and $\gamma$.}
\label{DOS_w0gam}
\end{figure}
\begin{figure}
\includegraphics[clip=true,width=0.40\textwidth]{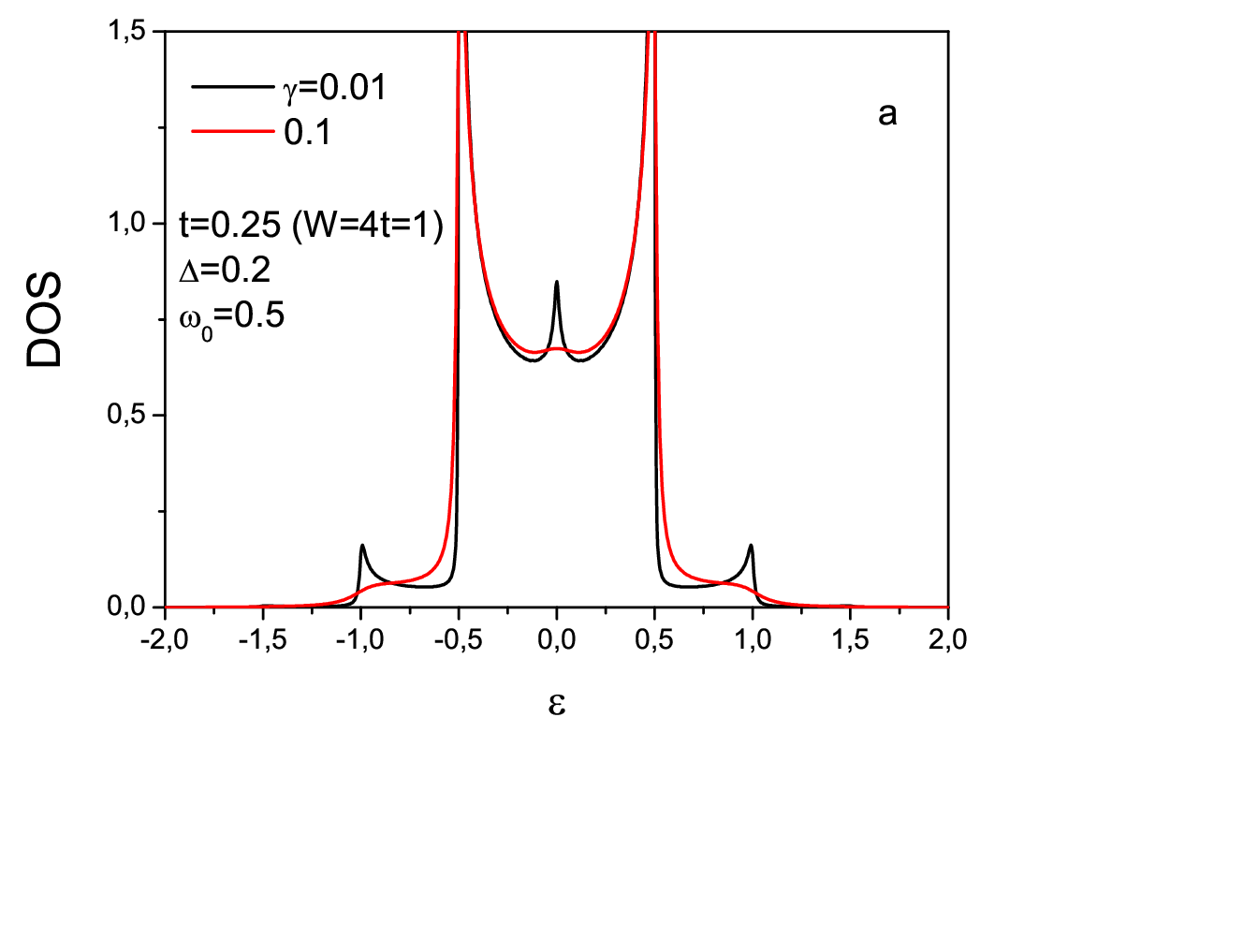}
\includegraphics[clip=true,width=0.40\textwidth]{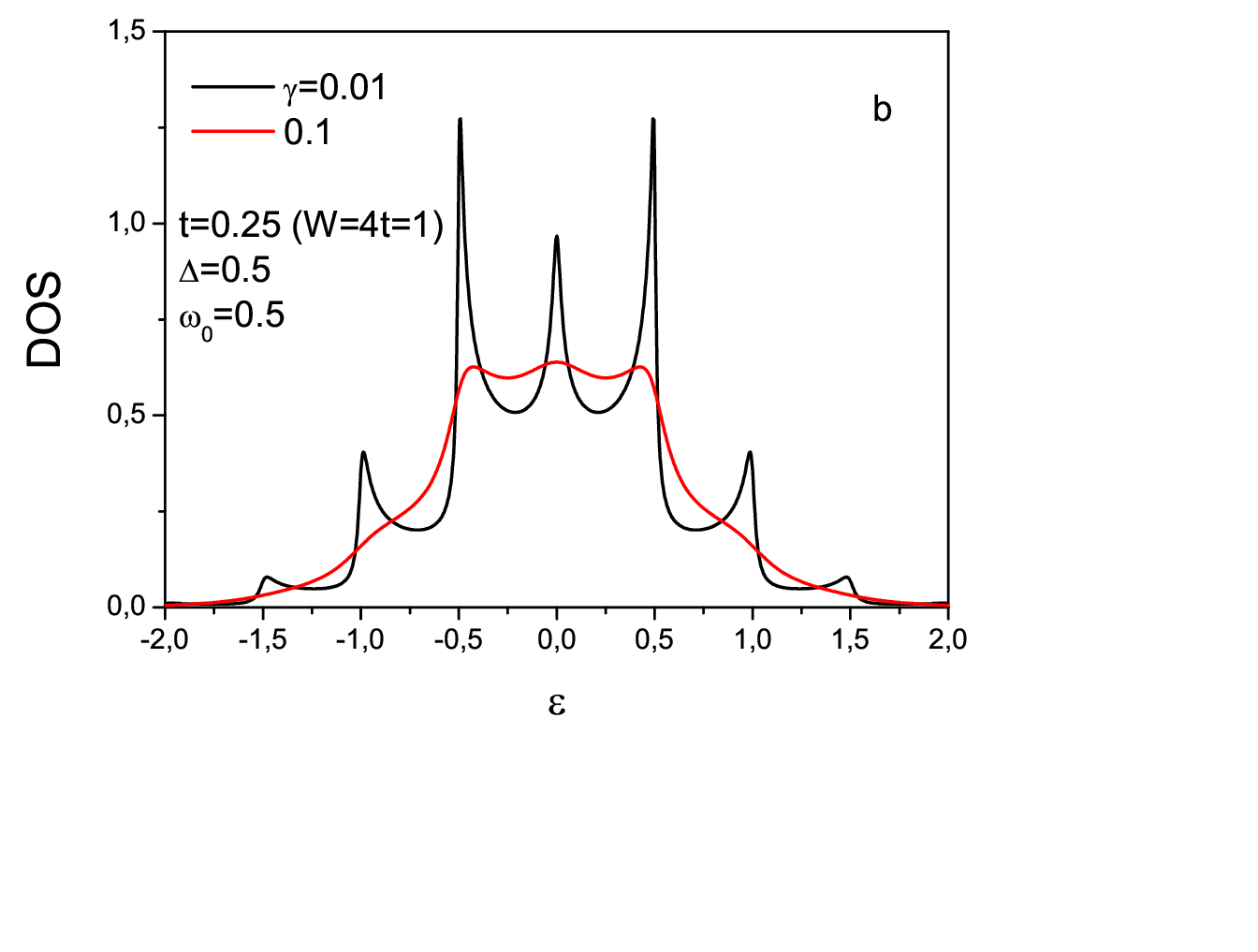}
\includegraphics[clip=true,width=0.40\textwidth]{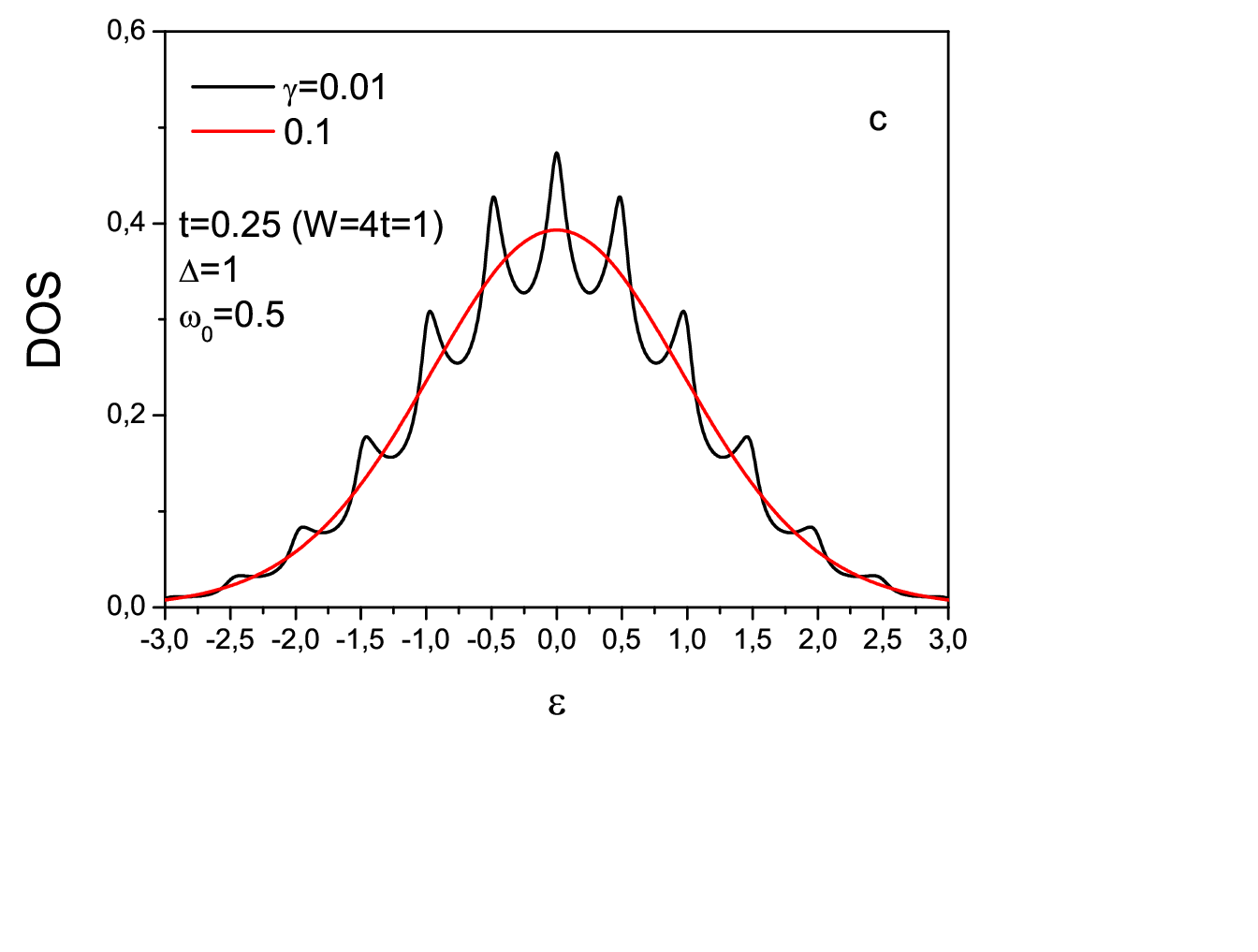}
\includegraphics[clip=true,width=0.40\textwidth]{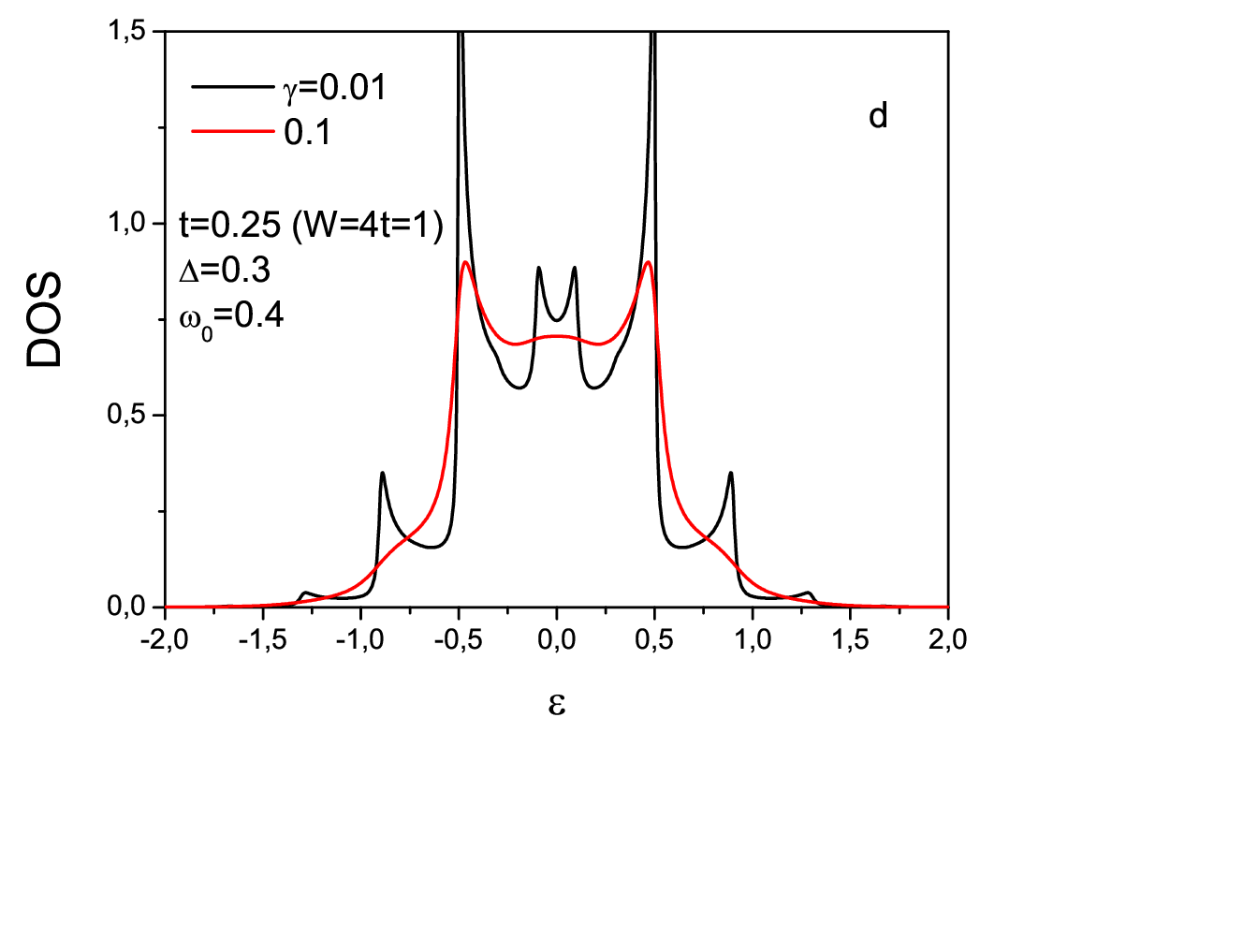}
\includegraphics[clip=true,width=0.40\textwidth]{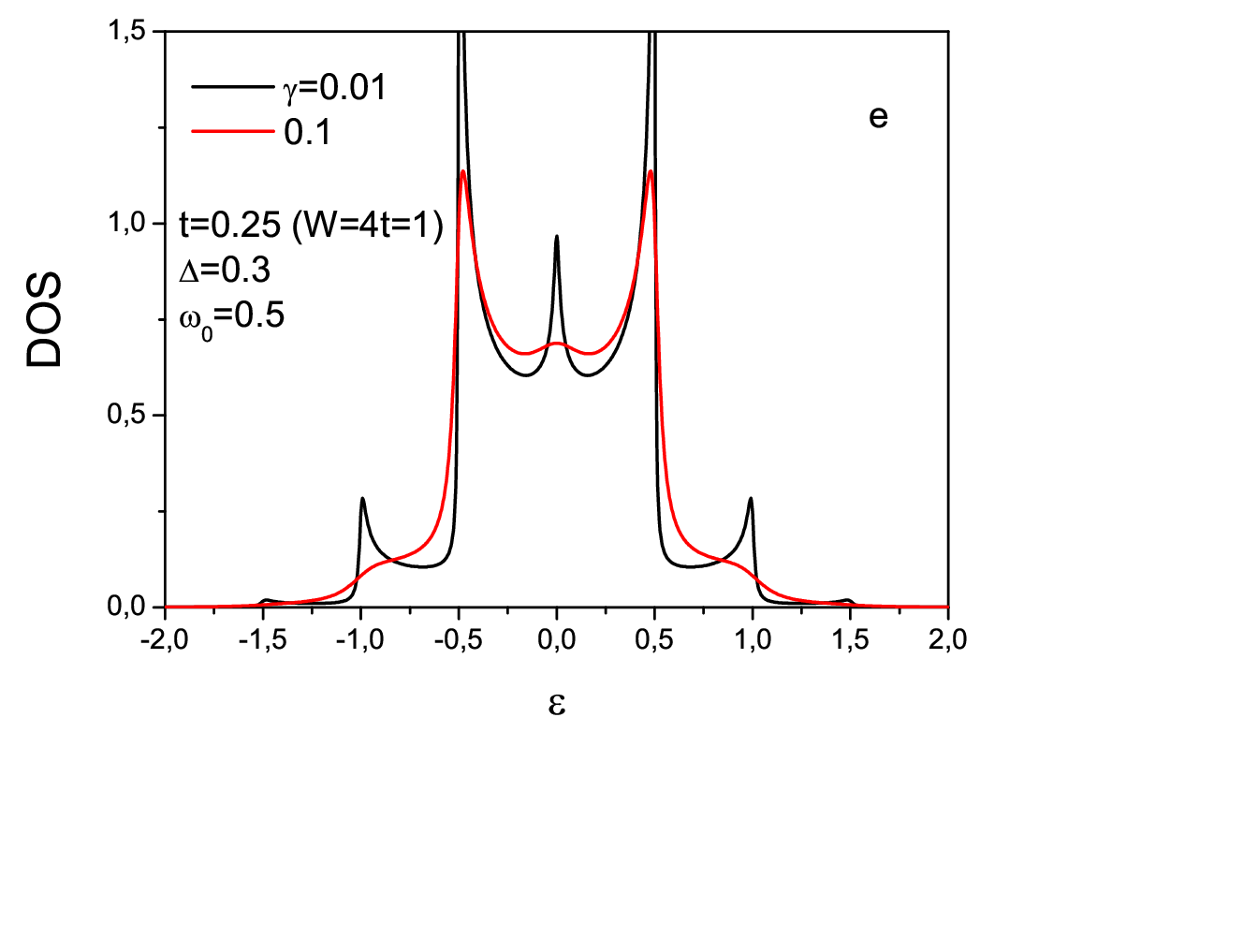}
\includegraphics[clip=true,width=0.40\textwidth]{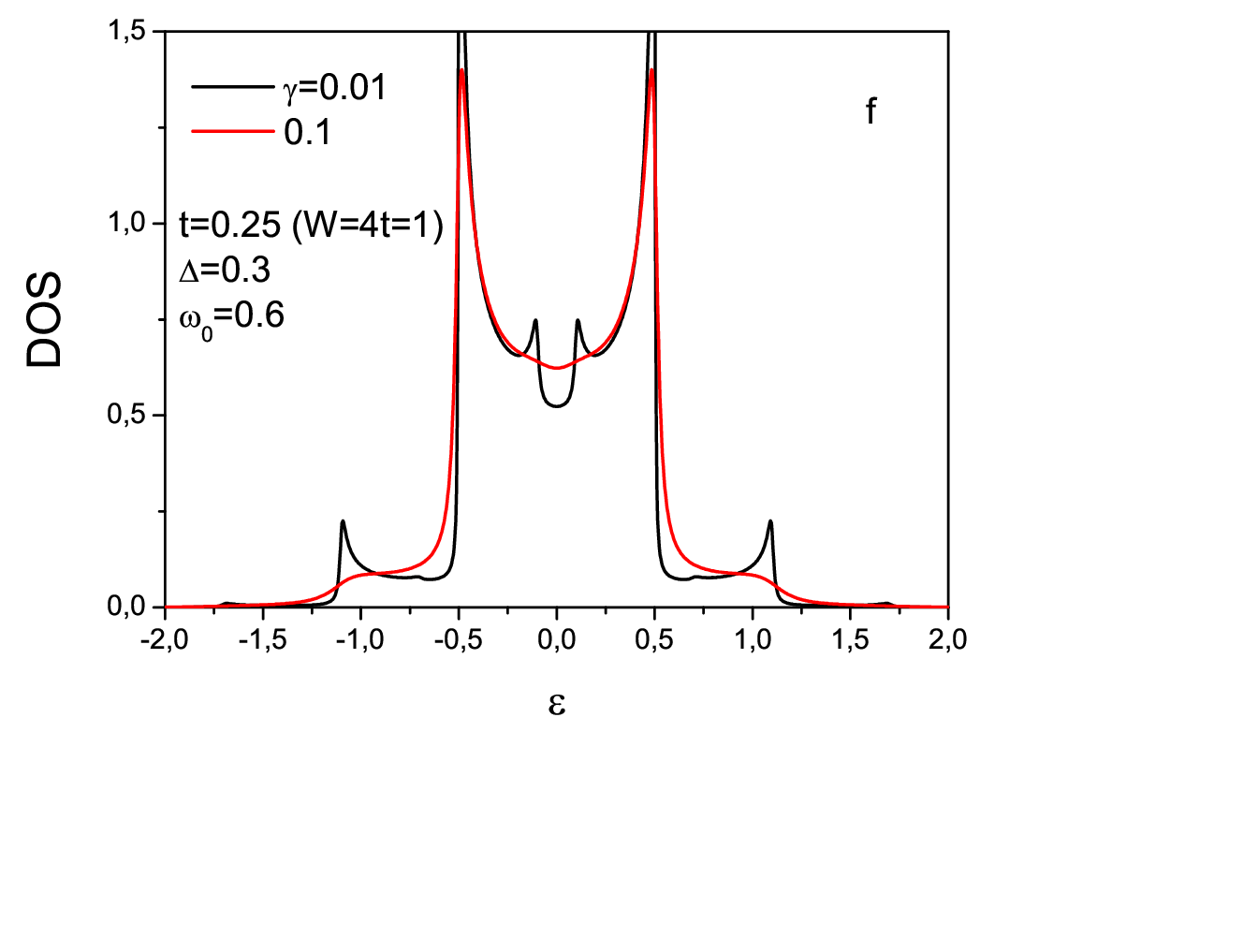}
\caption{Density of states for one -- dimensional chain
with initial bandwidth $W=4t=1$ for different $\Delta$, $\omega_0$
and $\gamma$.}
\label{DOSd1}
\end{figure}
\begin{figure}
\includegraphics[clip=true,width=0.40\textwidth]{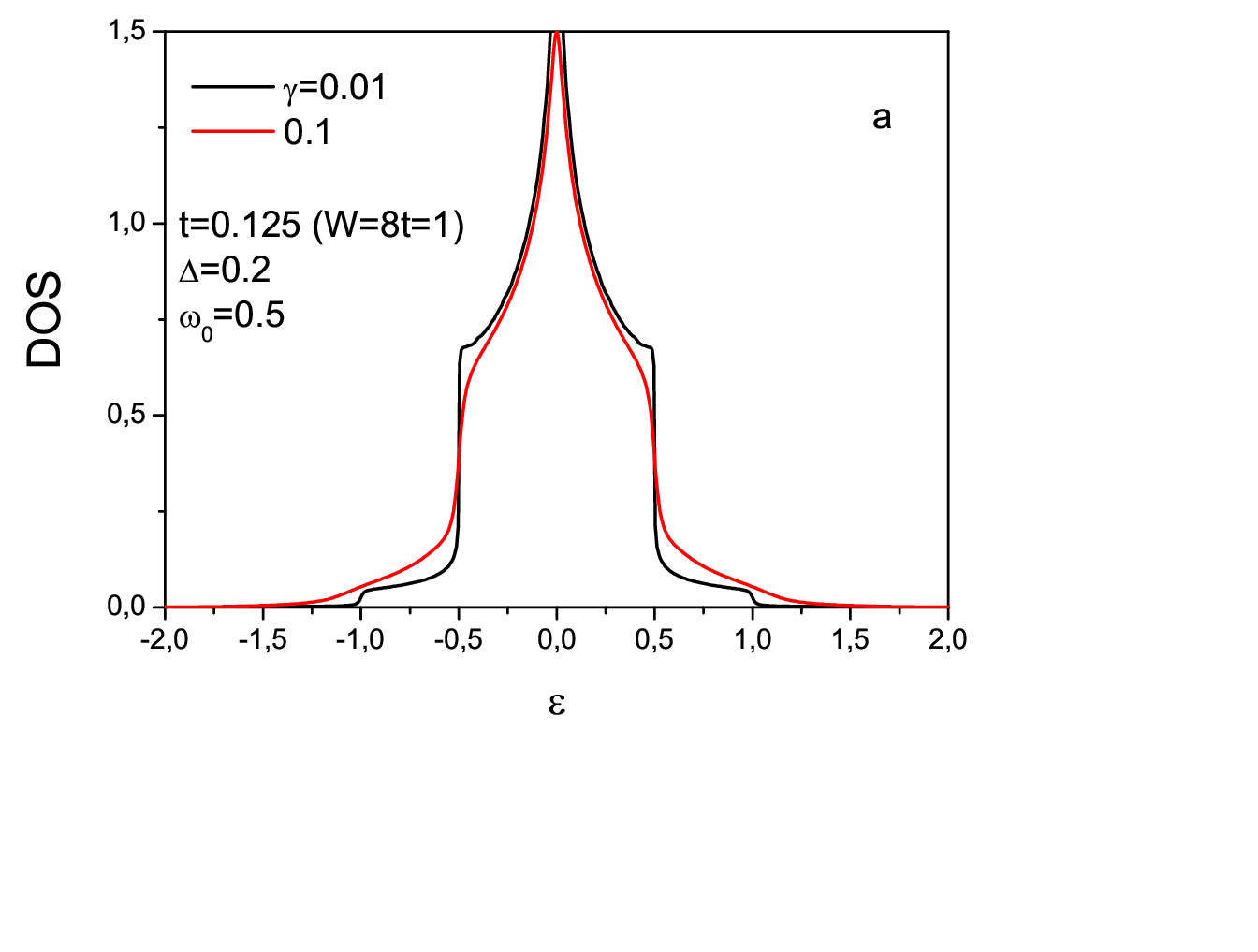}
\includegraphics[clip=true,width=0.40\textwidth]{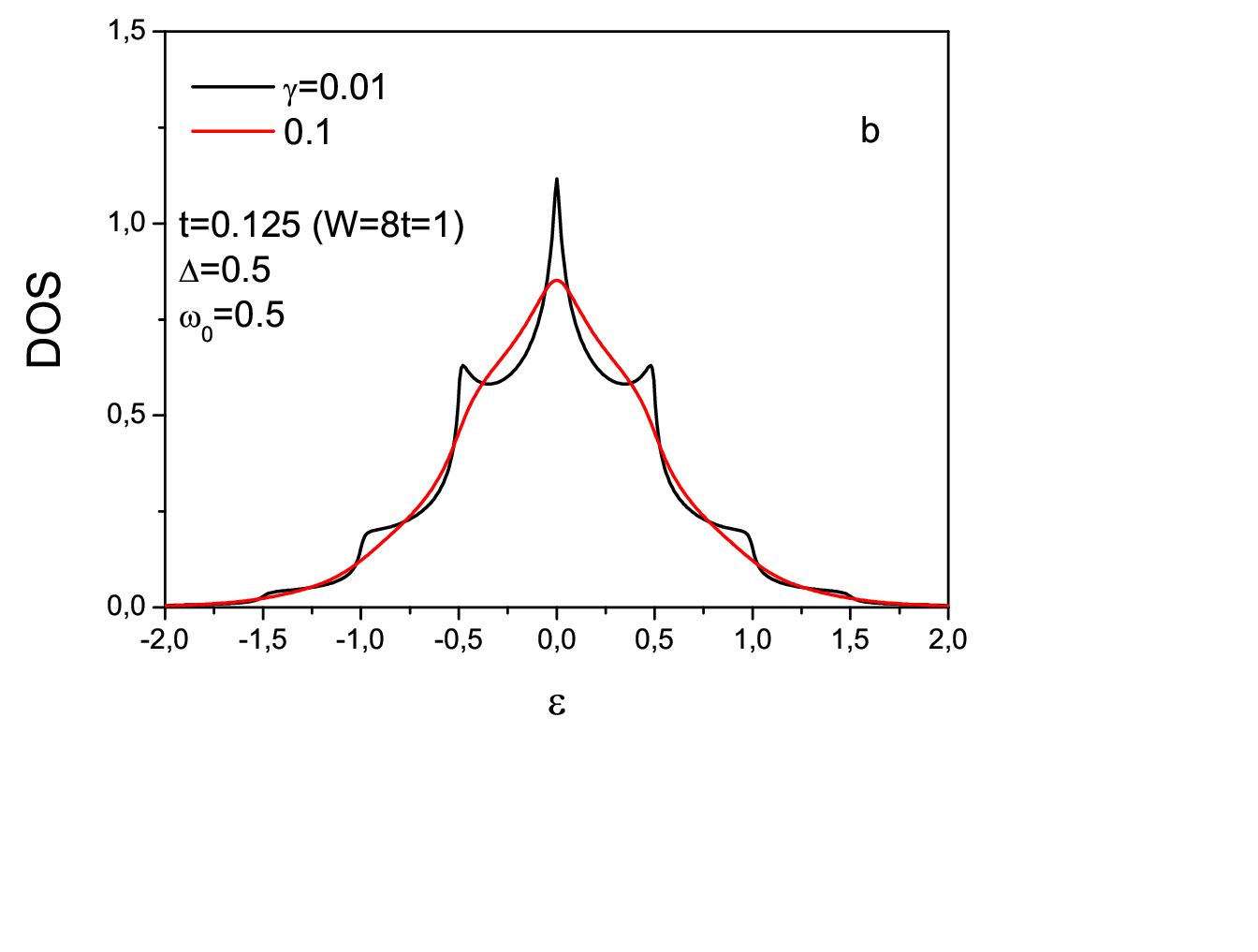}
\includegraphics[clip=true,width=0.40\textwidth]{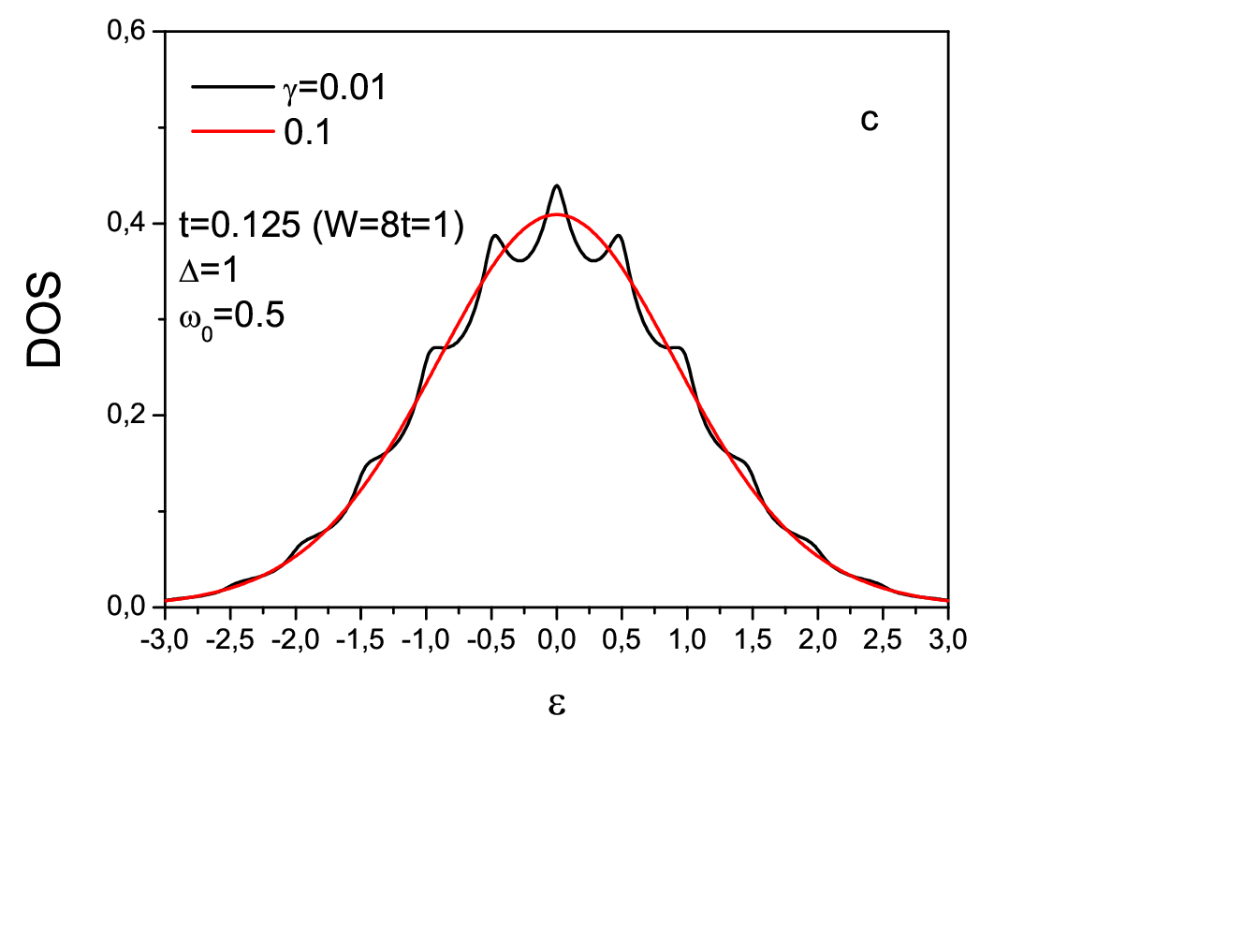}
\includegraphics[clip=true,width=0.40\textwidth]{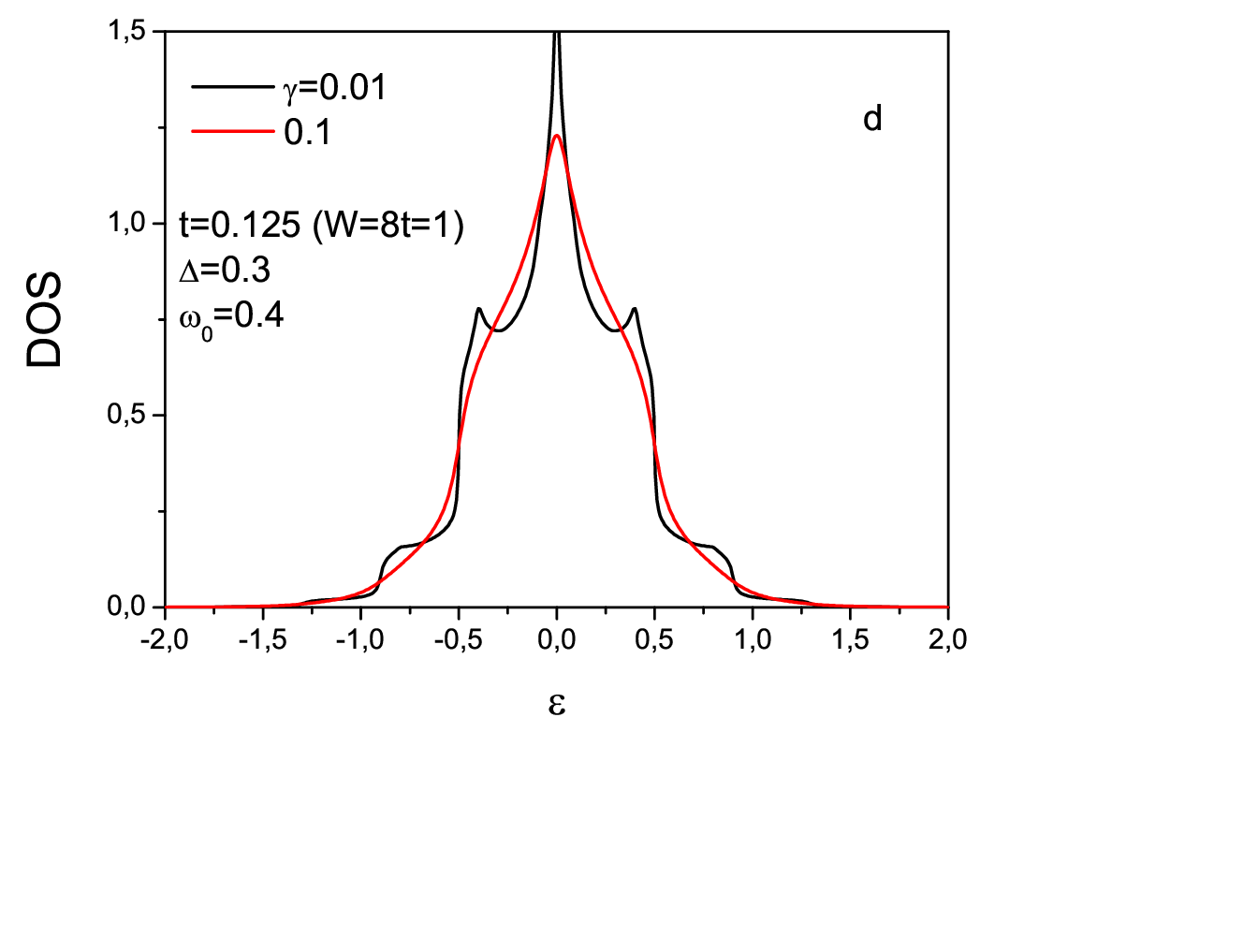}
\includegraphics[clip=true,width=0.40\textwidth]{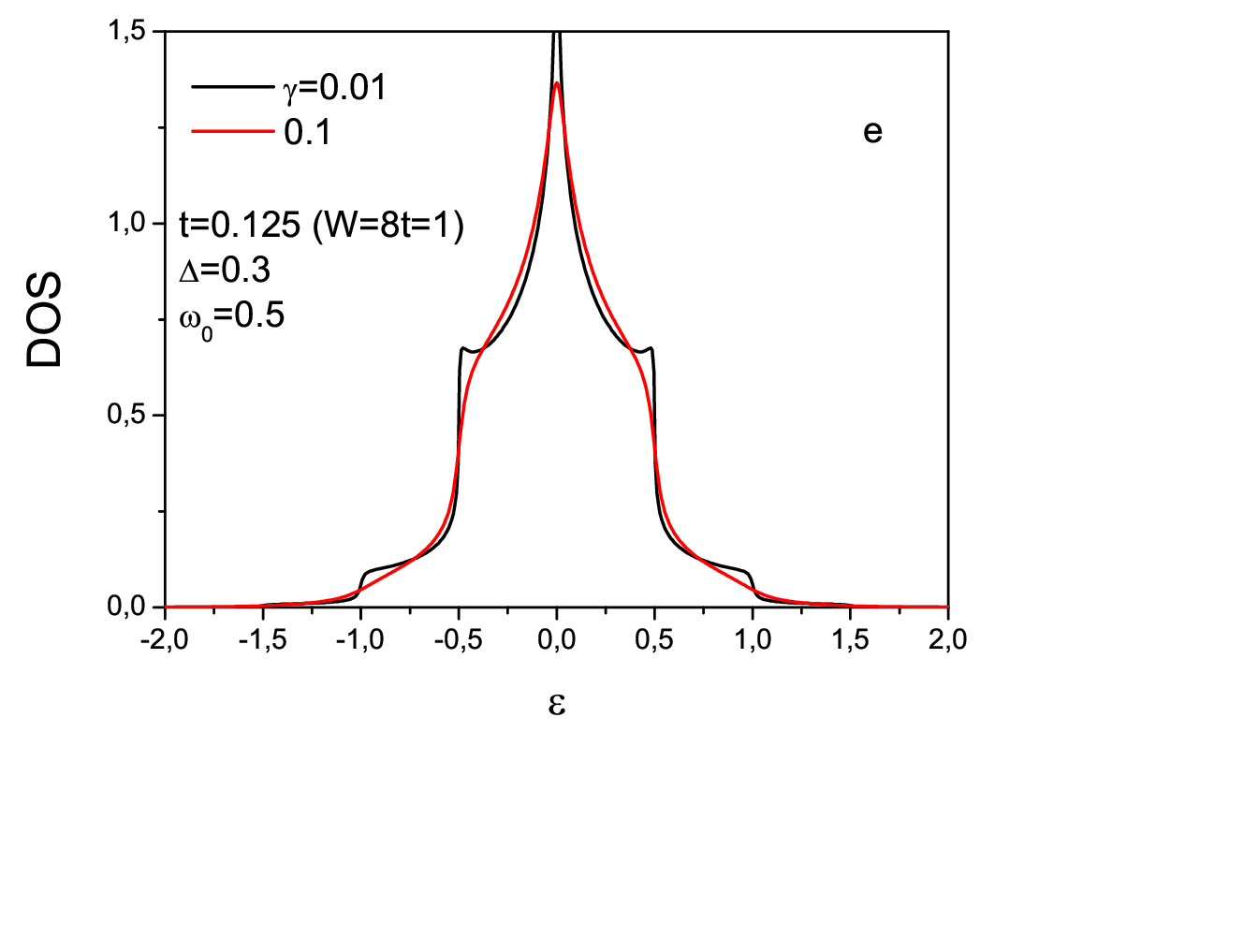}
\includegraphics[clip=true,width=0.40\textwidth]{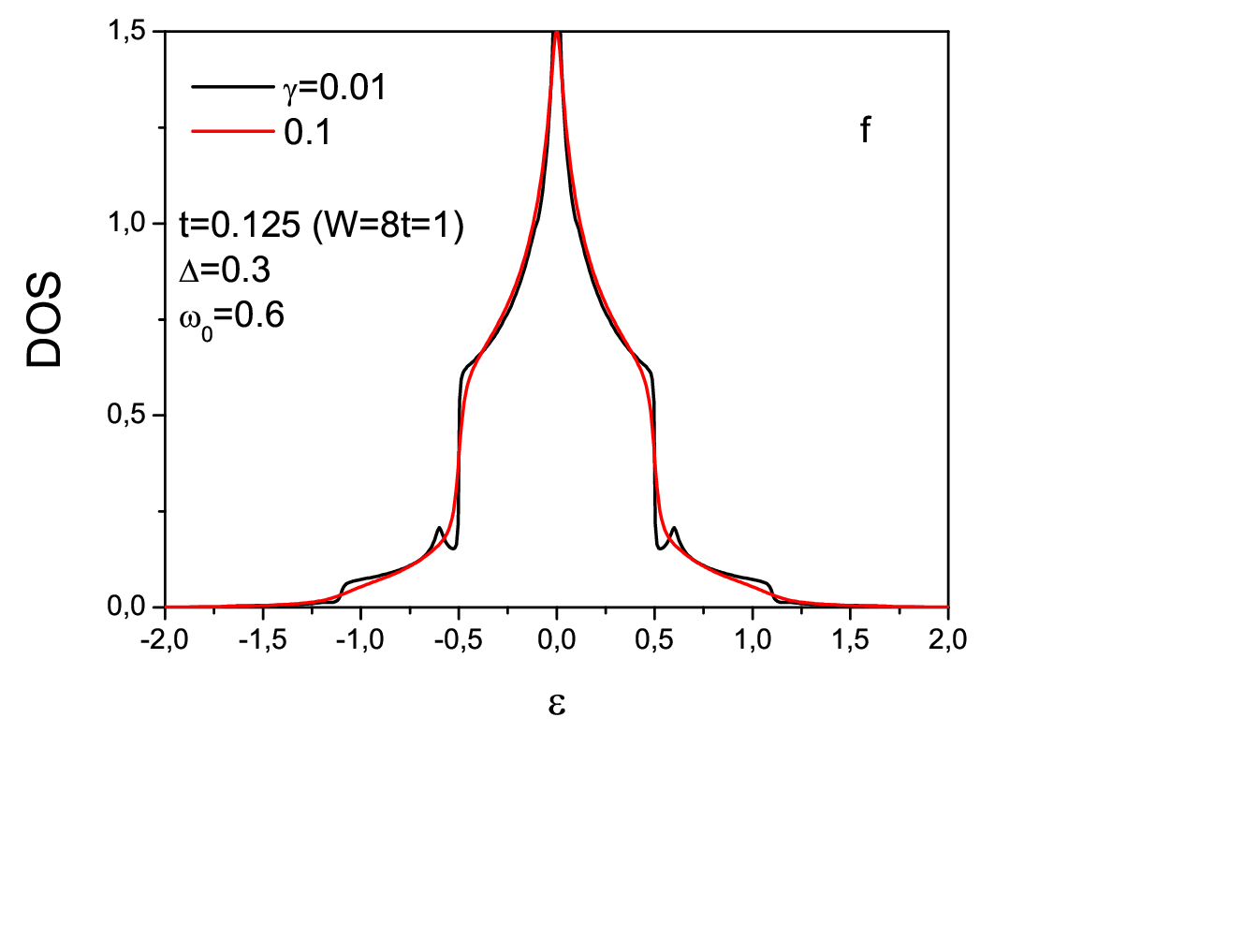}
\caption{Density of states in two -- dimensional lattice
with initial bandwidth $W=8t=1$ for different $\Delta$, $\omega_0$
and $\gamma$.}
\label{DOSd2}
\end{figure}
\begin{figure}
\includegraphics[clip=true,width=0.40\textwidth]{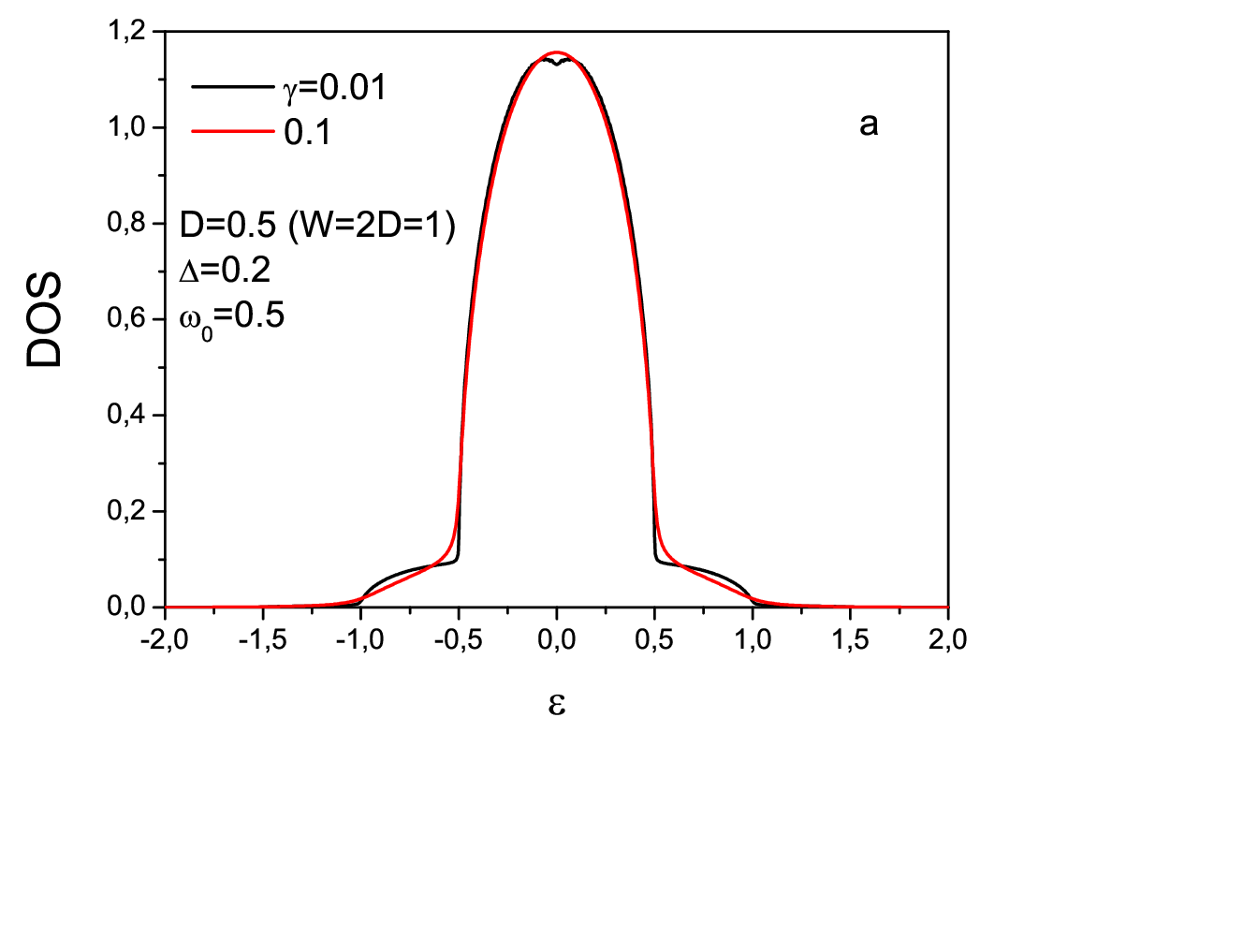}
\includegraphics[clip=true,width=0.40\textwidth]{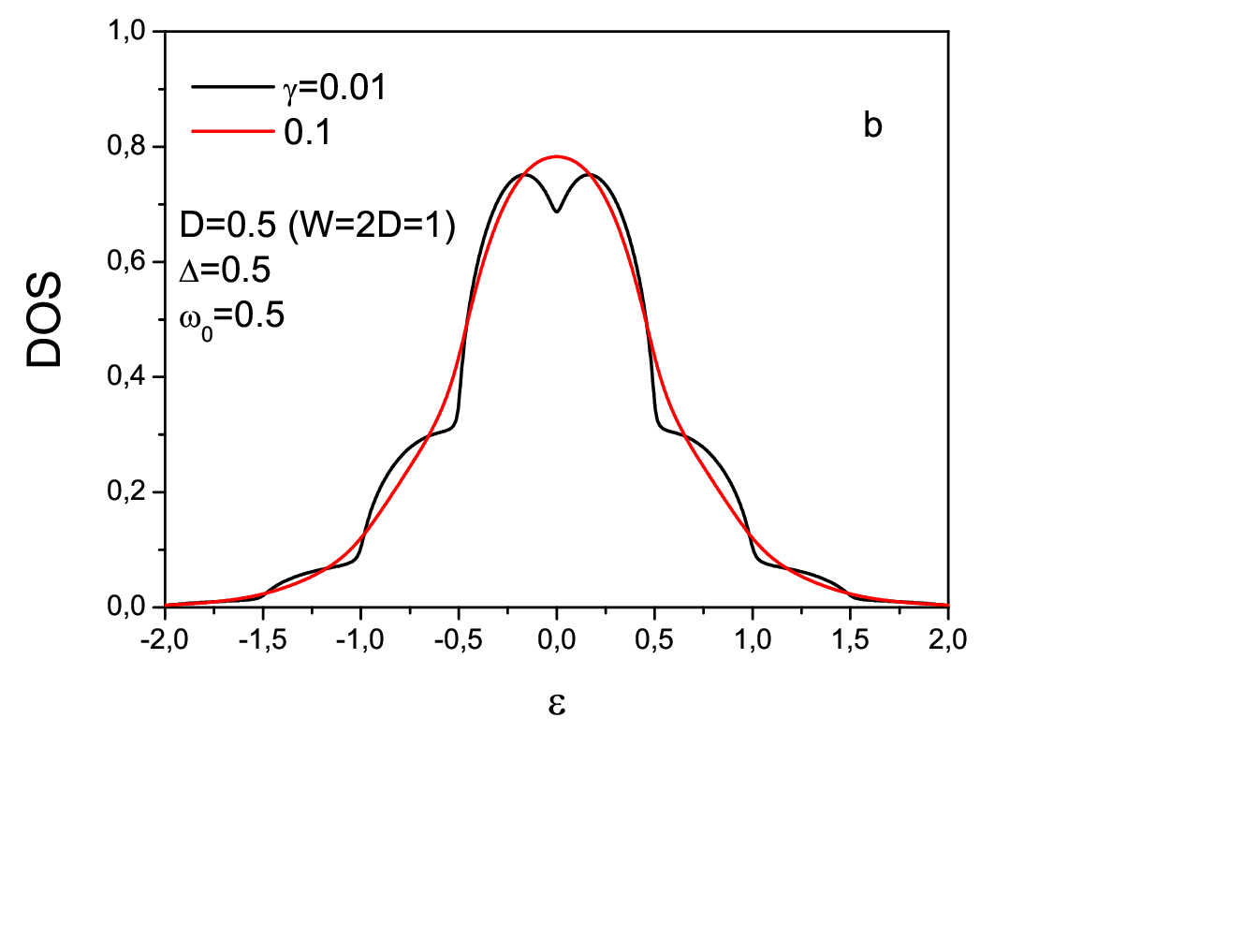}
\includegraphics[clip=true,width=0.40\textwidth]{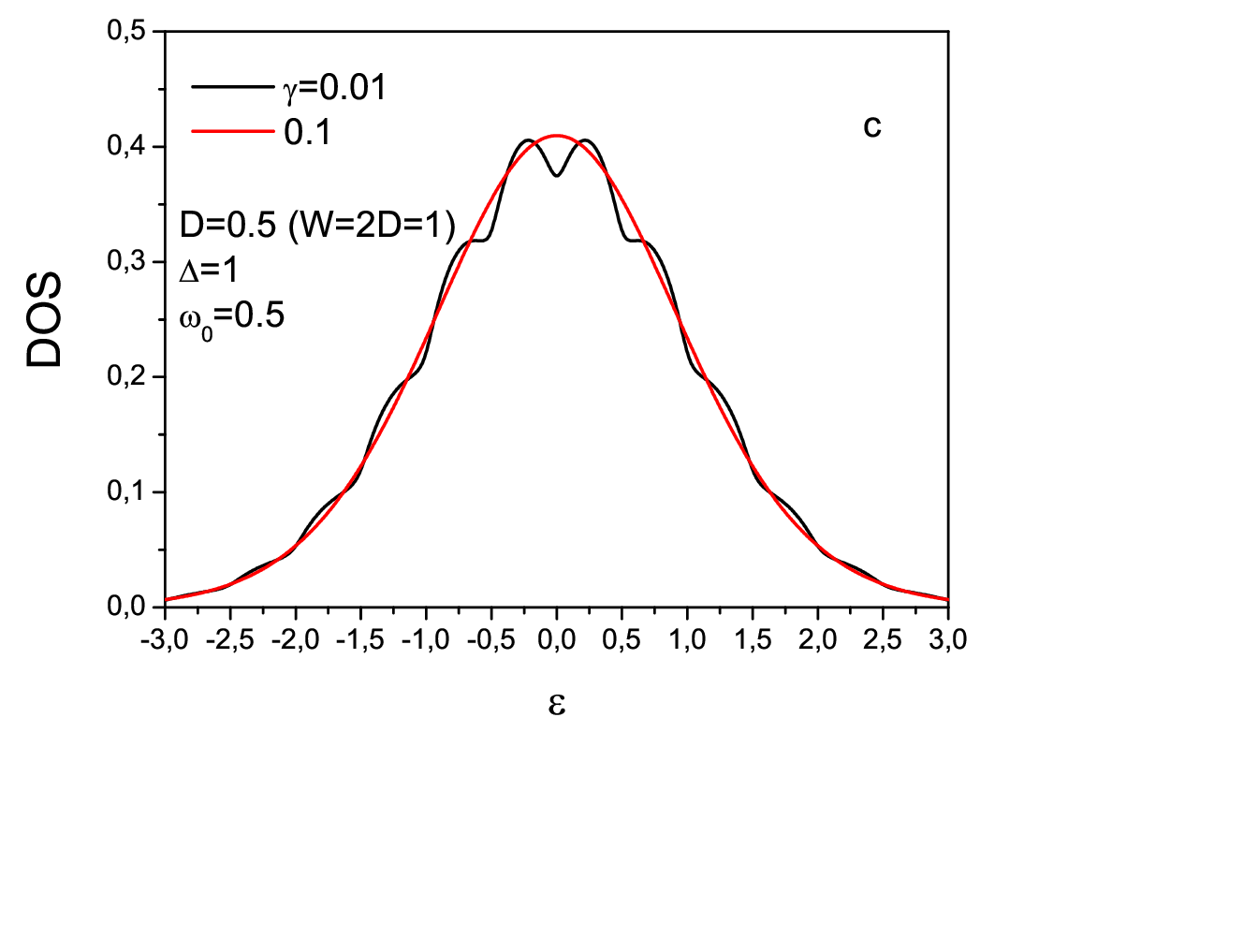}
\includegraphics[clip=true,width=0.40\textwidth]{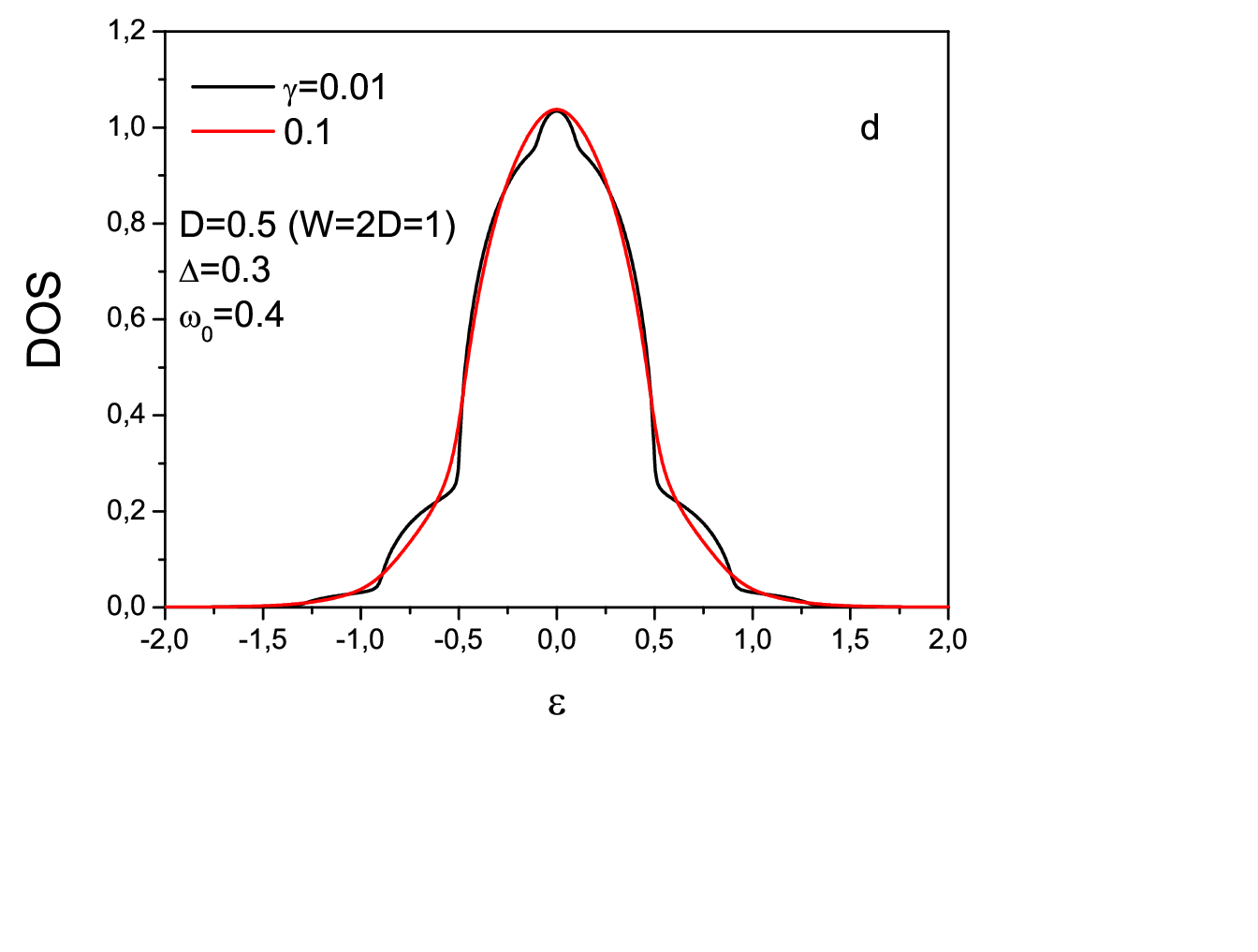}
\includegraphics[clip=true,width=0.40\textwidth]{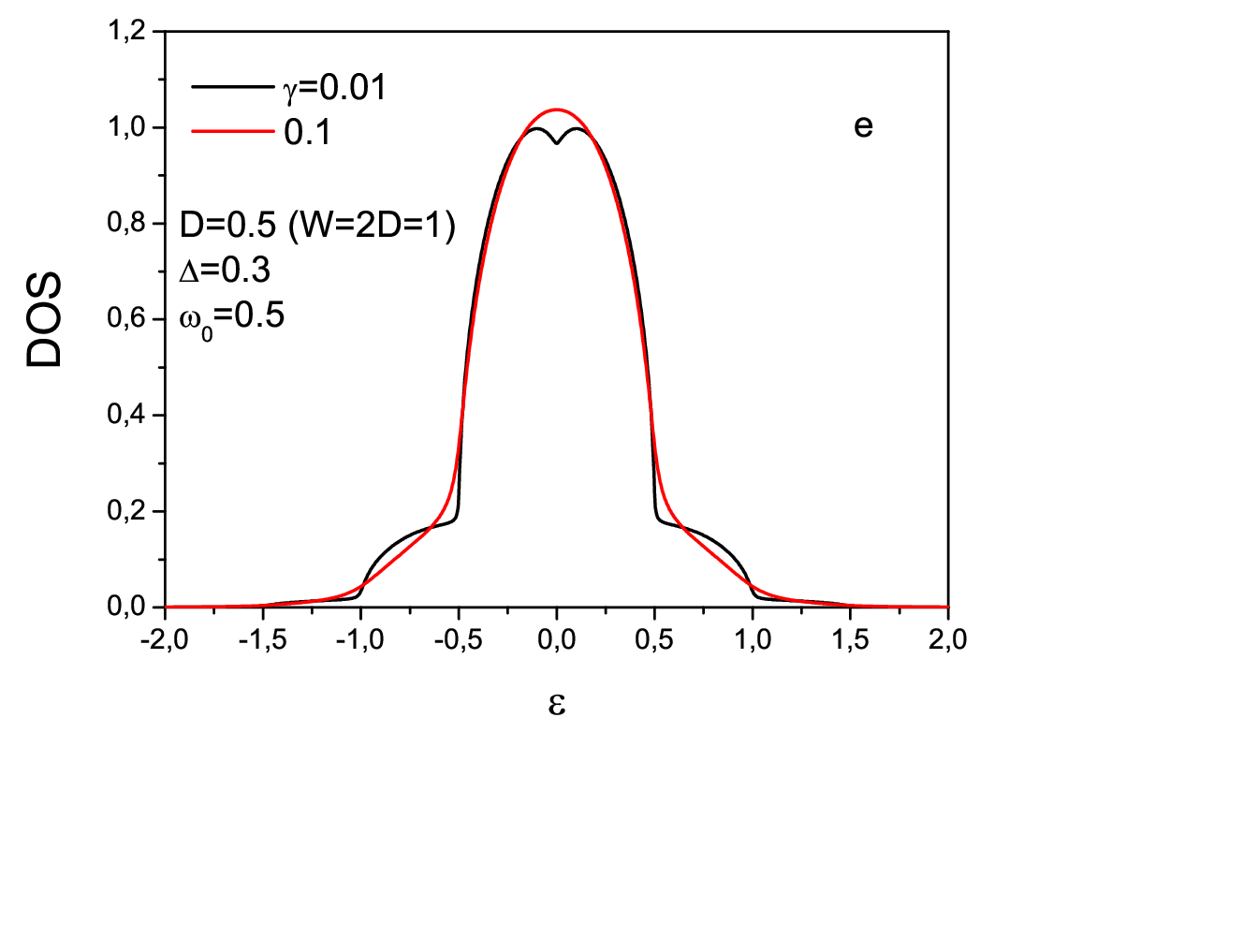}
\includegraphics[clip=true,width=0.40\textwidth]{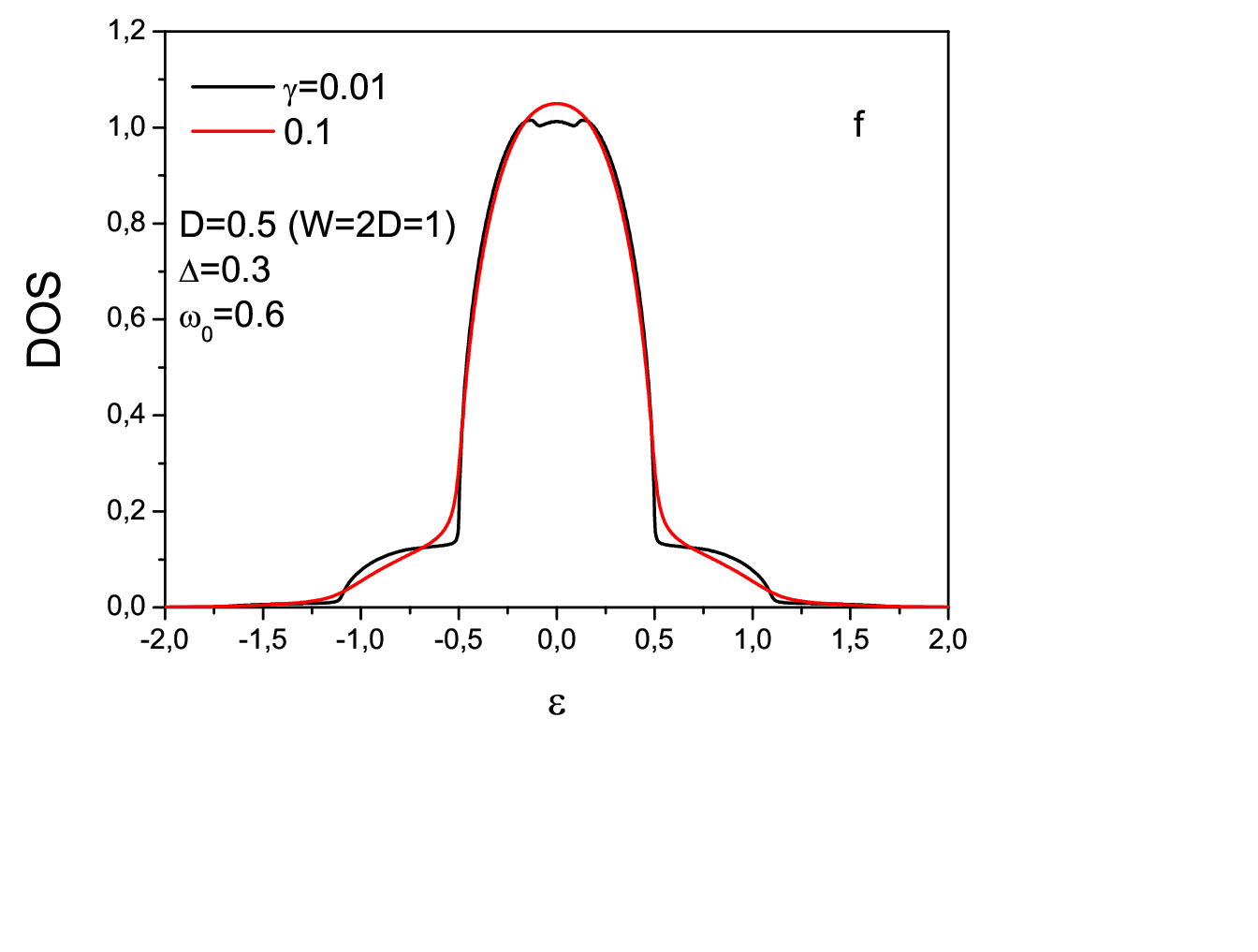}
\caption{Density of states of three -- dimensional system with
initial semi -- elliptic density of states with bandwidth $W=2D=1$
for different  $\Delta$, $\omega_0$ and $\gamma$.}
\label{DOSd3}
\end{figure}

In Fig. \ref{DOS_w0gam} we show spectral densities (densities of states in
quantum dot) in the model with finite transferred frequency for $\Delta=1$
and different values of  $\omega_0$ and $\gamma$.
We can see that in all cases for small $\gamma$ significant modulations of
the spectral density  appear with frequency $\omega_0$ with peaks of spectral
density appearing at energies $\epsilon=\pm n\omega_0$, where $n$ is integer.
The height of these peaks decreases with increasing $n$ and for
$\epsilon>3\Delta$ peaks are practically invisible. Increasing $\gamma$
leads to decreasing peak heights and starting from some values of $\gamma$
modulations with frequency $\omega_0$ become unobservable. Further increase of
$\gamma$ only somehow narrows Gaussian -- like spectral density,
as it was observed in Fig.\ref{DOSw00_gam} for the model with $\omega_0=0$.
At large enough values of $\gamma$, when no modulations of spectral density
with frequency $\omega_0$ are observed, the growth of $\omega_0$ only weakly
changes the spectral density (see Fig.\ref{DOS_w0gam}f) and we can use more
simple model with $\omega_0=0$. Note that the values of $\gamma$,
for which modulations of spectral density are observable depends on $\omega_0$.
In particular, for $\omega_0=0.1$ (Fig.\ref{DOS_w0gam}a) modulations are
observed only for $\gamma=0.0001$,
while for $\omega_0=0.5$ (Fig.\ref{DOS_w0gam}c) modulations are observable
already for $\gamma=0.05$.

As was already noted above it is not difficult to generalize our model to
consider not a single quantum well in dynamical random fields, but electron in
crystal lattice of  $d$ dimensions (in the following we take lattice parameter
$a\equiv 1$) with transfer integral between nearest neighbors $t$, which is
placed in a capacitor, with noise created at its plates, the same for all
lattice sites. This field is thus constant in space and the electron momentum
is not changed during scattering, so that the account of electron hops between
lattice sites is taken into account by a simple replacement
$\epsilon\to\epsilon-\epsilon_{\bf p}$, where $\epsilon_{\bf p}$ is band -- like
spectrum of electrons with quasimomentum ${\bf p}$.
In such a model the Green's function is given by:
\begin{equation}
G(\epsilon, {\bf p})=\int_{-\infty}^{\infty}d\epsilon '\frac{\rho(\epsilon ')}
{\epsilon-\epsilon_{\bf p}-\epsilon '+i\delta},
\label{analGp}
\end{equation}
where $\rho(\epsilon)$ is the spectral density (density of states) obtained
above for the problem of a single quantum dot.
Then for the density of states of our lattice model in $d$ dimensions in
dynamical random field we obtain:
\begin{equation}
N_d(\epsilon)=-\frac{1}{\pi}Im\sum_{\bf p}G(\epsilon, {\bf p})=
\int_{-\infty}^{\infty}d\xi N_{0d}(\xi)\rho(\epsilon-\xi),
\label{DOSd}
\end{equation}
where $N_{0d}(\xi)=\sum_{\bf p}\delta(\xi-\epsilon_{\bf p})$ is the ``bare''
density of states of $d$ dimensional system in the absence of random field.

For one -- dimensional chain:
\begin{equation}
\epsilon_p=-2t\cos(p)
\label{ep_d1}
\end{equation}
``Bare'' density of states in this case is:
\begin{equation}
N_{0d1}(\epsilon)=\frac{1}{\pi}\frac{1}{\sqrt{4t^2-\epsilon^2}}
\label{DOS0d1}
\end{equation}
and diverges at the band edges. Full densities of states for this model for
initial band of the width $W=4t=1$ and different values of random field
parameters are shown in Fig. \ref{DOSd1}.

For two -- dimensional lattice:
\begin{equation}
\epsilon_{\bf p}=-2t(\cos(p_x)+\cos(p_y)).
\label{ep_d2}
\end{equation}
``Bare'' density of states in this case has step -- like behavior at the
band edges and logarithmic Van-Hove singularity at the band center.
Full densities of states obtained in this model for the band with initial
width $W=8t=1$ and different values of random field parameters are shown in
Fig.\ref{DOSd2}.

To analyze three -- dimensional case we use as the ``bare'' the model
semi -- elliptic density of states:
\begin{equation}
N_{0d3}(\epsilon)=\frac{2}{\pi D^2}\sqrt{D^2-\epsilon^2},
\label{DOS0d3}
\end{equation}
where $D$ is the band half -- width. This model guarantees the valid
$\sim\epsilon^{1/2}$ ``bare'' density of states behavior near the band edges
for $d=3$. Full densities of states in this model for initial bandwidth
$W=2D=1$ and different values of random field parameters are shown in
Fig.\ref{DOSd3}.

Thus in all these models for small values of $\gamma$ we can observe modulations
of the density of states with frequency $\omega_0$.
Increasing  $\gamma$ leads to sharp weakening of these modulations.
The growth of random field amplitude  $\Delta$
(Figs.\ref{DOSd1},\ref{DOSd2},\ref{DOSd3}a,b,c) leads to some increase of
modulations amplitude and weakening of singularities
(Van - Hove, at band edges etc..), related to the ``bare'' density of states.
For $\Delta=W$ (Figs.\ref{DOSd1},\ref{DOSd2},\ref{DOSd3}c) density of states
pratically ``forgets'' the bare one. Increase of spatial dimensionality $d$
leads to weakening of the modulations.

In one -- dimensional chain (Fig.\ref{DOSd1}) for $\omega_0=0.5$
peaks at $\epsilon=\pm\omega_0$ coincide with band -- edges, where the bare
density of states (\ref{DOS0d1}) diverges, while the peak at $\epsilon=0$
appears at the minimum of the bare density of states.
Thus the peaks at $\epsilon=\pm\omega_0$ are effectively increased and can
can become larger than the weakened peak at $\epsilon=0$ (Fig.\ref{DOSd1}a,b,e).
This mutual influence of divergence in the bare density of states at the
bad edges in one dimension and modulations with frequency $\omega_0$ leads to
significant changes if the amplitude and shape of central peak (at $\epsilon=0$)
with small changes of $\omega_0$ close to $\omega_0=0.5$ (Fig.\ref{DOSd1}d,e,f).

For two -- dimensional lattice Van - Hove divergence is at the band center,
and central peak of modulations is always significantly larger than peaks at
$\epsilon=\pm\omega_0$ and its shape is only weakly changes with small variations
of $\omega_0$ close to $\omega_0=0.5$ (Fig.\ref{DOSd2}d,e,f).

For three -- dimensional model modulations in the density of states with
frequency $\omega_0$ are weak enough and for $\omega_0=0.5$ even a small dip is
observed in the density of states in the middle of the band (at  $\epsilon=0$)
(Fig.\ref{DOSd3}a,b,c,e). Small variations of $\omega_0$ close to
$\omega_0=0.5$ significantly change the shape of this weak feature at the
band center (Fig.\ref{DOSd3}d,e,f).

\section{Conclusions}

Our analysis shows a plenty of new and interesting results, which can be
derived even for this simple enough version of the generalized dynamical Keldysh
model for the case of random fields with finite transferred frequency.
It seems obvious that this model can have a direct relation to situations
realized in real systems with quantum dots, which are used in different
microelectronic devices, while the frequency $\omega_0$ can be related to
the clock frequency of these devices. Of course, the current simplest model is
oversimplified, but one can hope that the results obtained can be useful also
for the analysis of processes in realistic devices.

The question of experimental realization of our model remains open.
In principle, the studies of quantum dots in the specially created
(e.g. by electrotechnical means) random field seems quite feasible, though
parameters of interaction with this are to be specially chosen to make the
results discussed above observable. All this is also directly related to
electronic systems (lattices) of different dimensionalities placed in a
random field created on ``capacitor'' plates.

In real physical systems dynamical random fields can be created e.g. by phonons
in the classical limit, when the temperature is much larger than the
characteristic frequency of these phonons $\omega_0$. For example, we can
consider electron scattering at the interface of metallic film and dielectric
substrate. It is well known that scattering with small transferred momenta
(almost ``forward'' scattering) can appear at the interface of metallic
monolayer of FeSe on the substrate made of ionic SrTiO$_3$ insulator
\cite{ARPES_FeSe_Nature}, which leads to interesting models of superconductivity
enhancement in this system \cite{FeSeSTO}. Unfortunately we can not apply the
analysis given above to this system, because the frequency of optical phonon in
SrTiO$_3$ is pretty high and it can not be considered as classical (external
random field). However, we can not exclude the existence of similar systems
(structures) with ``soft'' enough optical phonons.

As was already noted above, the model with a single quantum well is directly
generalized to the case of several wells \cite{KK,EK}, leading to Keldysh model
with multicomponent noise. In particular, the model with two wells is closely
related (in the variant with band electrons) to the exactly solvable model of
pseudogap state \cite{Sad1,Sad2,Won1,Won2,Sad3,SadTim}. Different models of
this kind were actively used to describe the pseudogap, appearing due to
electron scattering by fluctuations of short -- range order in one -- dimensional
models \cite{sad,Sad1,Sad2,Won1,Won2,Sad3,SadTim}, which were also generalized
for two -- dimensional case to describe pseudogap in high -- temperature
superconductors \cite{PS1,PS2,SK98,ufn1,Tamm}. In most of these papers only
scattering by quasi static fluctuations was considered. It is of great interest
to generalize these models for the case of dynamical fluctuations with finite
transferred frequency, created by appropriate ``soft'' modes. However, it is
clear that analysis of such models requires further development of 
methods used in this paper. We hope to perform such studies in some future.


\end{document}